\documentclass{aa}
\usepackage[breaklinks=true]{hyperref}
\newcommand{\orcid}[1]{\protect\href{https://orcid.org/#1}{\protect\includegraphics[width=8pt]{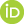}}} 

\usepackage{graphicx}
\usepackage{ragged2e}
\usepackage{txfonts}
\usepackage{supertabular}
\begin{document} 

\title{Observations of type Ia supernova SN\,2020nlb up to 600\,days\\after explosion, and the distance to M85
}

\titlerunning{Observations of SN\,2020nlb}
\authorrunning{S.~C. Williams et al.}
   \author{
          S.~C.~Williams\inst{1,2}\orcid{0000-0001-8178-0202}
          \and
          R.~Kotak\inst{2}\orcid{0000-0001-5455-3653}
          \and
          P.~Lundqvist\inst{3,4}\orcid{0000-0002-3664-8082}
          \and
          S.~Mattila\inst{2,5}\orcid{0000-0001-7497-2994}
          \and
          P.~A.~Mazzali\inst{6,7}\orcid{0000-0001-6876-8284}
          \and
          A.~Pastorello\inst{8}\orcid{0000-0002-7259-4624}
          \and
          A.~Reguitti\inst{9,8}\orcid{0000-0003-4254-2724}
          \and
          M.~D.~Stritzinger\inst{10}\orcid{0000-0002-5571-1833}
          \and
          A.~Fiore\inst{11, 8} \orcid{0000-0002-0403-3331}
          \and 
          I.~M.~Hook\inst{12}\orcid{0000-0002-2960-978X}
          \and
          S.~Moran\inst{2}\orcid{0000-0001-5221-0243}
          \and
          I.~Salmaso\inst{8,13}\orcid{0000-0001-5221-0243}
          }

    \institute{Finnish Centre for Astronomy with ESO (FINCA), Quantum, Vesilinnantie 5, University of Turku, FI-20014 Turku, Finland\\
    \email{steven.williams@utu.fi}
    \and
    Department of Physics and Astronomy, University of Turku, FI-20014 Turku, Finland
    \and
    Department of Astronomy, AlbaNova University Center, Stockholm University, SE-10691 Stockholm, Sweden
    \and
    The Oskar Klein Centre, AlbaNova, SE-10691 Stockholm, Sweden
    \and
    School of Sciences, European University Cyprus, Diogenes street, Engomi, 1516 Nicosia, Cyprus
    \and
    Astrophysics Research Institute, Liverpool John Moores University, IC2, Liverpool Science Park, 146 Brownlow Hill, Liverpool L3 5RF, UK
    \and
    Max-Planck Institute for Astrophysics, Karl-Schwarzschild-Straße 1, 85748 Garching, Germany
    \and
    INAF - Osservatorio Astronomico di Padova, Vicolo dell’Osservatorio 5, I-35122 Padova, Italy
    \and
    INAF - Osservatorio Astronomico di Brera, Via E. Bianchi 46, I-23807, Merate (LC), Italy
    \and
    Department of Physics and Astronomy, Aarhus University, Ny Munkegade 120, DK-8000 Aarhus C, Denmark
    \and
    Institut f\"ur Theoretische Physik, Goethe Universit\"at, Max-von-Laue-Str. 1, 60438 Frankfurt am Main, Germany
    \and
    Department of Physics, Lancaster University, Lancaster, Lancashire LA1 4YB, UK
    \and
    Dipartimento di Fisica e Astronomia ``G. Galilei'', Università degli Studi di Padova, Vicolo dell’Osservatorio 3, 35122 Padova, Italy
    }

   \date{Received ???; accepted ???}
 
  \abstract
   {The type Ia supernova (SN~Ia) SN\,2020nlb was discovered in the Virgo Cluster galaxy M85 shortly after explosion. Here we present observations that include one of the earliest high-quality spectra and some of the earliest multi-colour photometry of a SN~Ia to date. We calculated that SN\,2020nlb faded $1.28\pm0.02$\,mag in the \textit{B} band in the first 15\,d after maximum brightness. We independently fitted a power-law rise to the early flux in each filter, and found that the optical filters all give a consistent first light date estimate. In contrast to the earliest spectra of SN\,2011fe, those of SN\,2020nlb show strong absorption features from singly ionised metals, including Fe~{\sc ii} and Ti~{\sc ii}, indicating lower-excitation ejecta at the earliest times. These earliest spectra show some similarities to maximum-light spectra of 1991bg-like SNe~Ia. The spectra of SN\,2020nlb then evolve to become hotter and more similar to SN\,2011fe as it brightens towards peak. We also obtained a sequence of nebular spectra that extend up to 594\,days after maximum light, a phase out to which SNe~Ia are rarely followed. The [Fe~{\sc iii}]/[Fe~{\sc ii}] flux ratio (as measured from emission lines in the optical spectra) begins to fall around 300 days after peak; by the $+594$\,d spectrum, the ionisation balance of the emitting region of the ejecta has shifted dramatically, with [Fe~{\sc iii}] by then being completely absent. The final spectrum is almost identical to SN\,2011fe at a similar epoch. Comparing our data to other SN~Ia nebular spectra, there is a possible trend where SNe that were more luminous at peak tend to have a higher [Fe~{\sc iii}]/[Fe~{\sc ii}] flux ratio in the nebular phase, but there is a notable outlier in SN\,2003hv. 
   Finally, using light-curve fitting on our data, we estimate the distance modulus for M85 to be $\mu_0=30.99\pm0.19$\,mag, corresponding to a distance of $15.8^{+1.4}_{-1.3}$\,Mpc.
   }


   \maketitle

\section{Introduction}

\begin{figure*}
\centering
\includegraphics[width=2\columnwidth]{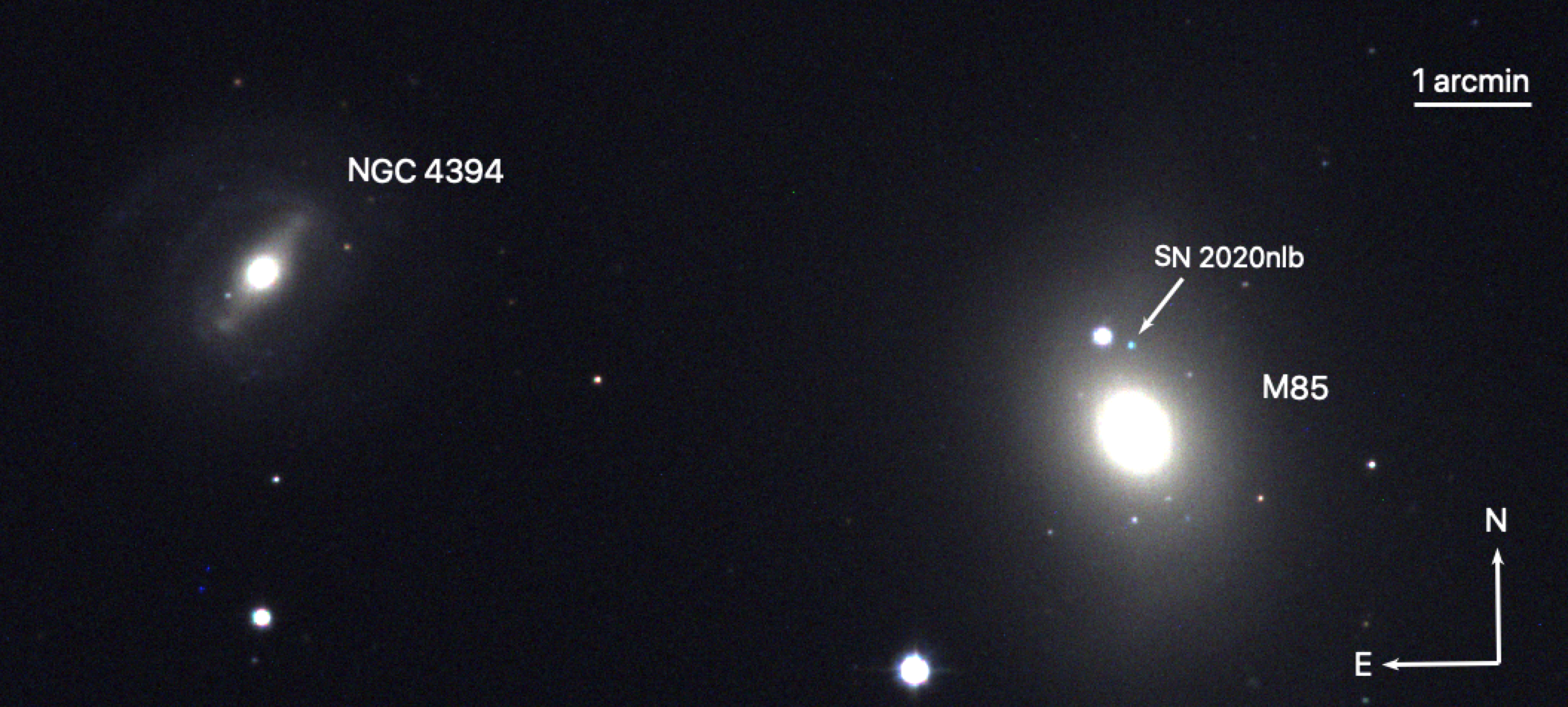}
\caption{\textit{BVr} image of SN\,2020nlb in its environment. The data were taken with the 67/91 Schmidt Telescope on 2021 Jan 11, when the SN was 183\,days after maximum light. By then, the SN was entering the nebular phase and had strong emission around the \textit{B} and \textit{V} band, hence the blue colour that the SN exhibits in this image.
}
\label{finder}
\end{figure*}

Type Ia supernovae (SNe~Ia) are generally accepted to be thermonuclear explosions of carbon-oxygen white dwarfs (CO WDs; \citealp{1960ApJ...132..565H,2000ARA&A..38..191H,2011Natur.480..344N}). Their light curves are almost entirely powered by the radioactive decay of $^{56}$Ni synthesised in the explosion, as it first decays to $^{56}$Co and then to stable $^{56}$Fe \citep{1962PhDT........25P,1969ApJ...157..623C,1982ApJ...253..785A}. Using a simple parameterisation of the width and colour of their light curves, SNe~Ia are standardisable candles, making them valuable distance indicators \citep{1993ApJ...413L.105P,1996ApJ...473...88R,1998A&A...331..815T}. This, combined with their high peak optical luminosities ($M_B\sim-19.3$\,mag), make them important cosmological probes \citep{1998AJ....116.1009R,1999ApJ...517..565P,2018ApJ...859..101S}. 

While there is a consensus over the type of primary star that produces a SN~Ia explosion, a CO WD, the configuration of the progenitor stellar system is still unclear. As isolated CO WDs cannot spontaneously explode, all SNe~Ia are thought to originate from binary (or higher) systems. The most plausible progenitor models for producing a significant fraction of SNe~Ia can be broadly classified based on the nature of the companion star to the exploding CO WD, as single-degenerate (a binary of a WD and a non-degenerate companion star; \citealp{1973ApJ...186.1007W}) or as double-degenerate (a WD--WD system; \citealp{1984ApJS...54..335I,1984ApJ...277..355W}). Whether normal SNe~Ia are produced by the explosion of a Chandrasekhar-mass ($M_{\mathrm{Ch}}$; 1.4\,$M_{\odot}$) WD or a sub-$M_{\mathrm{Ch}}$ WD is also debated. Models for both types of explosion are plausibly able to reproduce many of the observables seen in SNe~Ia, including the range of peak brightnesses that SNe~Ia show (e.g.\ \citealp{1995ApJ...444..831H}, \citealp{2010ApJ...714L..52S}, \citealp{2013MNRAS.429.2127B}, \citealp{2017hsn..book..317T}, \citealp{2018ApJ...854...52S}, \citealp{2019ApJ...873...84P}).

Type Ia supernovae spectra are characterised by strong absorption features from intermediate-mass elements (IMEs) such as silicon and sulphur prior to and near maximum light, which over time increasingly give way to a spectrum dominated by features associated with iron-group elements (IGEs). The ejecta of normal SNe~Ia are stratified, with the inner layers being dominated by IGEs, outside of which is an IME-dominated region, and the very outer layers are abundant in oxygen and any unburnt carbon. However, this is only a broad overall picture, and the spectroscopic evolution of SNe~Ia suggests there is substantial mixing between the different burning products (e.g.\ \citealp{1985ApJ...294..619B,2005MNRAS.360.1231S,2007Sci...315..825M,2011MNRAS.410.1725T}). In order to study the outer, higher-velocity regions of the ejecta, spectroscopy at the earliest times is required.

The advent of deep all-sky surveys such as the Asteroid Terrestrial-impact Last Alert System (ATLAS; \citealp{2018PASP..130f4505T}) and Zwicky Transient Facility \citep{2019PASP..131a8002B} has made it possible to study SNe~Ia shortly after explosion in greater numbers. Over the last few years, a number of SNe~Ia have been observed to exhibit rises that do not conform to the traditional power-law rise \citep[e.g.][]{2015Natur.521..328C,2017ApJ...845L..11H,2019ApJ...870L...1D,2019ApJ...870...13S,2020ApJ...898...56M}. In some cases, the early excess flux is sufficiently high for the SNe to display a bump feature in the early light curve (e.g.\ SN\,2017cbv, \citealp{2017ApJ...845L..11H}; SN\,2019yvq, \citealp{2020ApJ...898...56M}). There are a number of potential causes of such a feature, for example: companion star interaction \citep{2010ApJ...708.1025K}, circumstellar medium (CSM) interaction \citep{2016ApJ...826...96P}, and `excess' radioactive isotopes in the outer ejecta \citep{2017MNRAS.472.2787N,2019ApJ...873...84P,2020arXiv200702101M}. Even if no bump is seen in the light curve, the early spectra and light curves could potentially, at least in principle, distinguish between explosion scenarios (see e.g.\ \citealp{2014MNRAS.441..532D,2017MNRAS.472.2787N}). The early-time luminosity and spectral features are sensitive to how the $^{56}$Ni is distributed throughout the ejecta.

SN\,2020nlb was discovered on 2020 Jun 25.25\,UT by ATLAS at an AB magnitude in the ATLAS-orange ($o$) filter of $m_{o}$ = $17.44\pm0.08$ \citep{2020TNSAN.126....1T}. The pre-discovery non-detection from ATLAS was $>$\,19.7\,mag (with the ATLAS-cyan filter) on 2020 Jun 23.28\,UT. We estimate in this work that first light occurred at 2020 Jun $23.1\pm0.1$ UT. We spectroscopically confirmed the transient as a SN~Ia 0.7\,days after discovery \citep{2020TNSAN.127....1F,2020TNSCR1934....1W}. \citet{2021ApJ...922...21S} used X-ray observations to place an upper limit on the pre-explosion mass loss of $<$\,$9.7\times10^{-9}$\,M$_{\odot}$\,yr$^{-1}$. They also searched for hydrogen or helium emission in the nebular spectrum and placed upper limits on the mass of H or He swept up from any potential non-degenerate companion star of $M_{\mathrm{H}}\lesssim0.7-2\times10^{-3}$\,M$_{\odot}$ and $M_{\mathrm{He}}\lesssim4\times10^{-3}$\,M$_{\odot}$.

SN\,2020nlb exploded in M85 (NGC~4382), a peculiar S0 galaxy \citep{1991rc3..book.....D} in the Virgo Cluster. The primary distance indicator for nearby galaxies are Cepheid variables; however, these are not found in passive environments. Therefore, for early-type galaxies, an alternative must be used. The distance to M85 has been estimated using a number of different methodologies. Surface brightness fluctuation (SBF) measurements indicate the distance modulus, $\mu_0=31.26\pm0.05$\,mag \citep{2007ApJ...655..144M}. Using the planetary nebula luminosity function (PNLF), a distance modulus of $30.79\pm0.06$\,mag was derived by \citet{1990ApJ...356..332J}. Discrepancies between PNLF and SBF distance moduli are common, and PNLF distance moduli are systematically $\sim$0.3\,mag lower than those from the SBF method \citep{2012Ap&SS.341..151C}.

In this work, we present our observations of SN\,2020nlb, beginning 0.7\,d after discovery. This includes one of the earliest ever spectra and some of the earliest multi-colour ($u'BVg'r'i'$) photometric data of a SN~Ia. Our first set of observations end 33\,d after peak brightness, when the SN became unobservable due to its proximity to the Sun. When it became observable again, we resumed follow-up and obtained a sequence of nebular spectra, with the final observations taken 594\,days after peak brightness. Given that M85 has no distance estimate using either Cepheid variables or the tip of the red giant branch, we also used our light curve of SN\,2020nlb to estimate the distance to the galaxy.

\section{Data}
\subsection{Photometry}
We obtained \textit{$u'$BV$g'r'i'$} follow-up using the IO:O \citep{smith_steele_2017} imager on the 2\,m Liverpool Telescope (LT; \citealp{2004SPIE.5489..679S}) and the Alhambra Faint Object Spectrograph and Camera (ALFOSC) on the 2.56\,m Nordic Optical Telescope (NOT), using the equivalent filter set. The IO:O imaging was reduced using the automatic LT IO:O pipeline and ALFOSC images were reduced using the \texttt{ALFOSCGUI} pipeline\footnote{ALFOSCGUI is a graphic user interface aimed at extracting SN spectroscopy and photometry obtained with FOSC-like instruments. It was developed by E.~Cappellaro. A description can be found at: \url{http://sngroup.oapd.inaf.it/foscgui.html}}. We also obtained some epochs of imaging with the 67/91 Schmidt Telescope.

All of our SN\,2020nlb photometry was calculated using PSF-fitting in \texttt{daophot} within \texttt{IRAF}\footnote{IRAF is distributed by the National Optical Astronomy Observatory, which is operated by the Association of Universities for Research in Astronomy (AURA) under a cooperative agreement with the National Science Foundation.} \citep{1986SPIE..627..733T}. The $g'r'i'V$ photometry of SN\,2020nlb was calibrated using a local sequence of stars from Pan-STARRS1 (PS1; \citealp{2016arXiv161205560C}). The \textit{V}-band magnitudes of the stars were computed from the PS1 magnitudes using the transformations in \citet{2012ApJ...750...99T}. The $u'B$ photometry was calibrated against stars in SDSS DR12 \citep{2015ApJS..219...12A}, with the \textit{B}-band magnitudes of the calibration stars computed using the transformations from \citet{2006A&A...460..339J} and the $u$ and $g$-band magnitudes from the SDSS catalogue. The errors on the measured SN magnitudes were generally dominated by the uncertainties on calculating the zero point of a given epoch (as determined from the local standard stars). For a few epochs where the signal-to-noise (S/N) of the SN was very low, we employed forced photometry at the location (using \texttt{daophot}). Our photometry of SN\,2020nlb is given in Table~\ref{tab:phot}.

\subsection{Spectroscopy}

We obtained a sequence of spectra using ALFOSC on the NOT as part of the NOT Unbiased Transient Survey (NUTS)\footnote{\url{https://sngroup.oapd.inaf.it/nuts.html}}. All spectra were taken using grism~4 (3200--9600\,\AA), with either $1.0''$ or $1.3''$ slit. Grism~4 with a $1.0''$ slit gives a resolution of $R=360$ or 16\,\AA. The ALFOSC spectra were extracted using the \texttt{ALFOSCGUI} pipeline. The flux calibration was performed using standard routines in \texttt{IRAF}, against observations of a spectrophotometric standard taken using the same instrumentation.

We also obtained several spectra using the SPectrograph for the Rapid Acquisition of Transients (SPRAT; \citealp{2014SPIE.9147E..8HP}) on the LT, which has a resolution of $R=350$ and wavelength coverage of 4000--8000\,\AA. These spectra were reduced to extracted 1D spectra using the automated SPRAT pipeline. The spectra were then calibrated against observations of the standard star BD+33\,2642 \citep{1990AJ.....99.1621O} with the same instrument setups, taken between 2020 Jun 22 and Jul 4. These flux calibrations were performed using standard routines in \texttt{IRAF}. We also obtained four spectroscopic epochs using the Low Resolution Spectrograph (LRS) on the Galileo National Telescope (TNG) and a single very late spectroscopic observation using FOcal Reducer/low dispersion Spectrograph 2 (FORS2; \citealp{1998Msngr..94....1A}) on the Very Large Telescope (VLT). Both the LRS and FORS2 spectra were reduced using standard routines in \texttt{IRAF}. A log of our spectroscopic observations is shown in Table~\ref{tab:log}.

\begin{figure}
\centering
\includegraphics[width=\columnwidth]{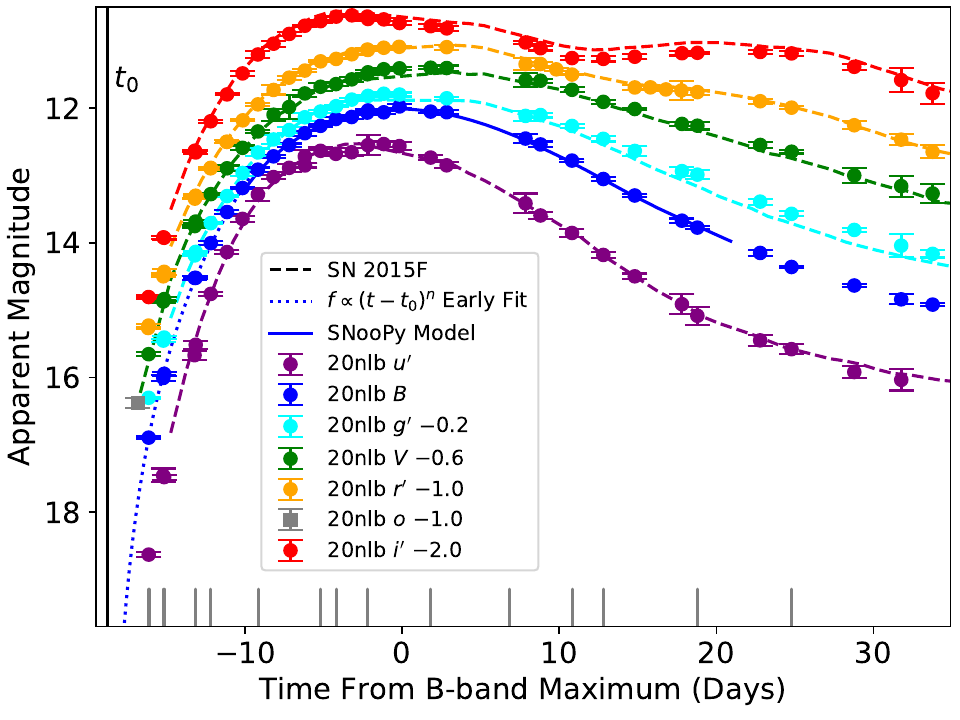}
\includegraphics[width=\columnwidth]{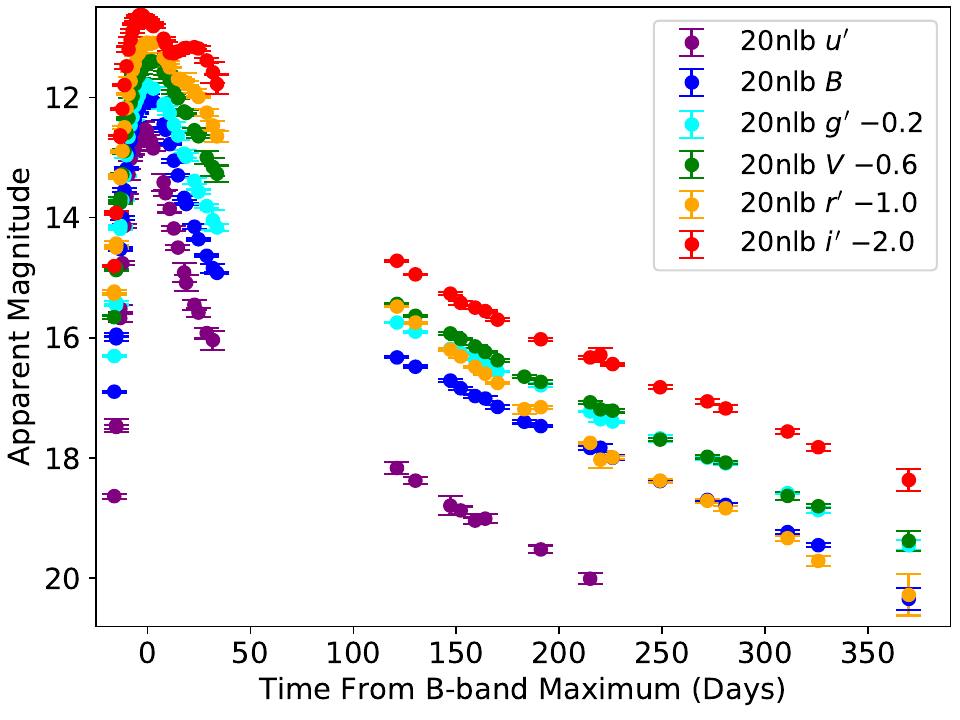}
\caption{Light curves of SN\,2020nlb. Top: Our early multi-colour light curves of SN\,2020nlb. The best-fit \texttt{SNooPy} model, and the best-fit $f\propto(t-t_0)^n$ to the early data, are shown for the \textit{B} band, using a solid and dotted line respectively. The ATLAS-\textit{o} discovery magnitude is also shown. The epochs that our spectra were taken are indicated by the grey lines at the bottom of the plot. The fitted first light time, $t_0$, is also indicated. The light curve of SN\,2015F (photometry from \citealp{2017MNRAS.464.4476C} and \citealp{2018ApJ...869...56B}) is shown for comparison with the dashed lines (except the \textit{B} band), which has been shifted to the distance modulus we derive for M85, with the distance modulus for SN\,2015F taken from \citet{2017MNRAS.464.4476C}. The light curves of both SNe have been corrected for foreground reddening only. \citet{2017MNRAS.464.4476C} estimate $E(B-V)_{\mathrm{host}}=0.085$ for SN\,2015F. Bottom: The full light curve of SN\,2020nlb stretching out to around one year after maximum light.
}
\label{lc}
\end{figure}

\section{Light curve}\label{sec:lcs}

Our multi-band light curves of SN\,2020nlb are shown in Fig.~\ref{lc}, where we also show photometry of SN\,2015F \citep{2017MNRAS.464.4476C,2018ApJ...869...56B}, which was a SN~Ia that we find to be spectroscopically and photometrically similar to SN\,2020nlb. Our light curves begin when the SN is 4.9\,mags below \textit{B}-band peak brightness, and $>$6 mags below $u'$-band peak brightness. At the time of the first $u'$-band measurement, the SN flux in that band was just 0.4\% of $u'$-band peak flux. We use the \texttt{SNooPy} SN~Ia light-curve-fitting code \citep{2011AJ....141...19B} on our \textit{B}-band photometry (using the `max$\_$model' fitting in \texttt{SNooPy}) to estimate the magnitude and time of peak \textit{B}-band brightness. For this and our subsequent light-curve fitting of SN\,2020nlb in this paper that utilises \texttt{SNooPy}, we use the filter responses of the various LT IO:O filters\footnote{\url{http://telescope.livjm.ac.uk/TelInst/Inst/IOO/}}. The \textit{B}-band peak occurred on MJD\,$59042.1\pm0.3$, with an apparent magnitude of $B=12.12\pm0.01$ (these values are very similar to those derived by \citealp{2021ApJ...922...21S}). We estimate the \textit{B}-band magnitudes that the SN faded during the first 15\,d from peak, $\Delta{m}_{15}(B)=1.28\pm0.02$\,mag. This makes it faster fading than the average of the normal SN~Ia distribution. The light curve implies a peak of $M_B=-19.1\pm0.2$\,mag, but note this is not an independent measurement, as the distance itself is derived from the light-curve fitting (see Section~\ref{dist}), and hence, in effect, the absolute peak brightness is implied from our measurements of the width and colour of the light curve. While using the distance derived from the light curve in the analysis of the SN is not ideal, we suggest that the parameters of this SN are in some tension with adopting the  SBF distance (see discussion in the last paragraph of Section~\ref{sec:ni}). We also note that the SBF and PNLF distances to the galaxy are in tension with each other, so the two would give very different SN luminosities to each other. For this work we adopt the SN-derived distance, but also note some of the key parameters for the case of the SBF distance.

\subsection{First light time and early rise}

\begin{figure}[ht]
\centering
\includegraphics[width=\columnwidth]{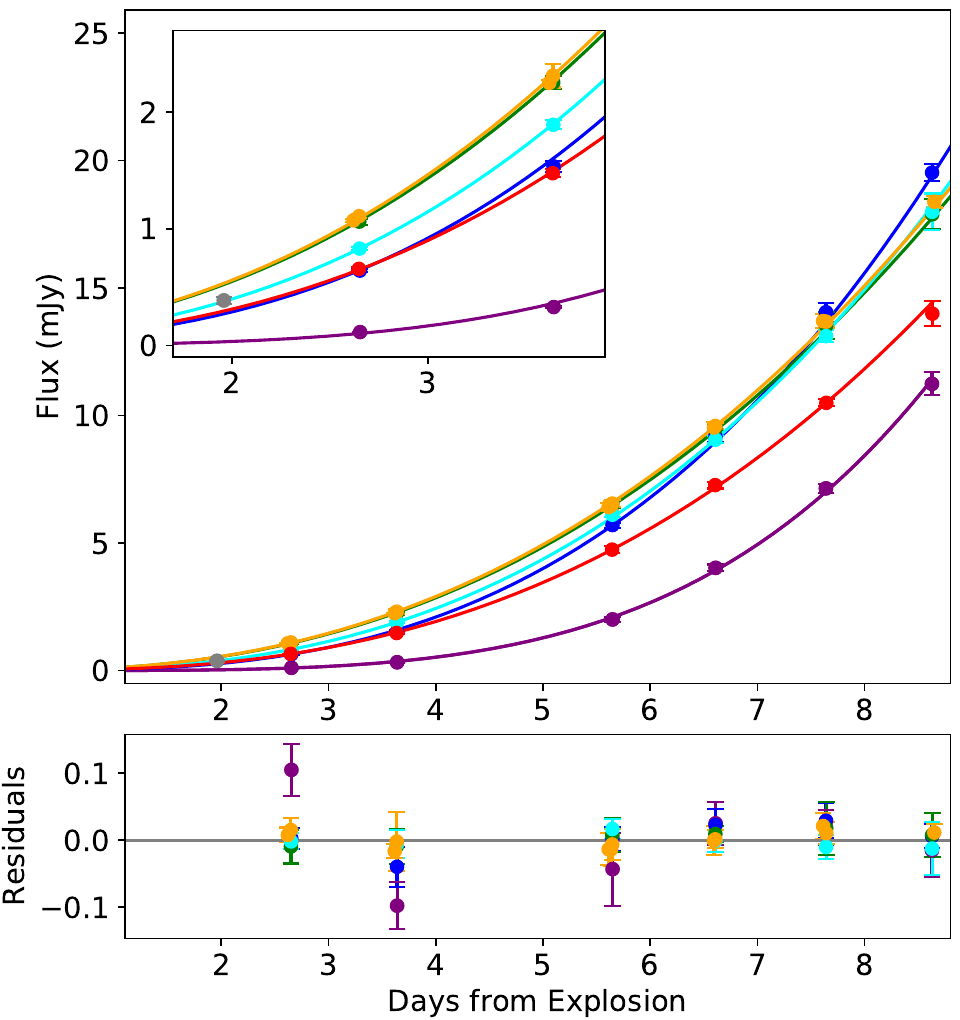}
\caption{Early light curve of SN\,2020nlb, showing the power-law fits with the first light date of $59023.3\pm0.1$\,MJD. The residuals (given as a proportion of the total flux at that epoch) of the fits are shown in the bottom panel.}
\label{lce}
\end{figure}

Given our early multi-colour coverage of this event, beginning 0.7\,d after the first detection, we are able to fit the early rise in each waveband. We fit a power law of the form $f\propto{(t+t_0)^n}$ to our early data (until the flux reaches 50\% of its maximum) in each filter, and the results for $t_0$ and $n$ in each case are listed in Table~\ref{tab:rise}. The best-fit values of $t_0$ in each filter are close to each other, with the exception of the $u'$-band, which also has a larger uncertainty. This is worth noting, as it indicates that for SNe~Ia, or at least those similar to SN\,2020nlb, the epoch of first light inferred from a power-law fit to early photometric data is largely independent of the filter used, at least in the optical. Therefore, for example, in the case of two SNe~Ia that did not have multi-colour early data, with perhaps only single-filter early observations, it would be acceptable to compare parameters that rely directly on $t_0$, without the concern of substantial systematic uncertainties arising from the difference in filters. The weighted mean first light time is $t_0=59023.3\pm0.1$\,MJD. We also re-fit the rise times, with a fixed $t_0=59023.3\pm0.1$ (with the uncertainty boot-strapped), and these values are also included in Table~\ref{tab:rise}. The fits are shown, along with our early photometry in Fig.~\ref{lce}. In the figure, we also display the ATLAS-$o$ discovery magnitude. The ATLAS-$o$ filter approximately covers the wavelengths of the \textit{r} and \textit{i} bands. It can be seen in the figure that the flux in the ATLAS-$o$ band at discovery lies between the \textit{r} and \textit{i}-band flux that would be predicted from the power-law fits to our data that begins 0.7\,d later. We note that the first light date we estimate here differs from that used by \citet{2021ApJ...922...21S}, who adopted the mid-point between the last non-detection and first detection as their explosion date.

\begin{table}[ht]
    \centering
    \caption{Fitted power-law rise in the form $f\propto(t-t_0)^n$, to our early data (when flux is less than half of the peak flux) of SN\,2020nlb in different filters. }
    \begin{tabular}{lccc}
    \hline\hline
        Band &$t_{0,\mathrm{band}}$ &\textit{n} ($t_{0,\mathrm{band}}$) &\textit{n} ($t_0$) \\
        \hline
         $u'$ &$59022.26\pm0.56$ &$4.87\pm0.47$ & $4.00\pm0.08$ \\
         $g'$ &$59023.44\pm0.19$ &$2.52\pm0.12$ & $2.62\pm0.05$\\
         $r'$ & $59023.12\pm0.13$ &$2.47\pm0.07$ & $2.38\pm0.04$ \\
         $i'$ & $59023.31\pm0.27$ &$2.61\pm0.16$ & $2.62\pm0.05$\\
         $B$ & $59023.37\pm0.17$ & $2.84\pm0.11$ & $2.89\pm0.05$ \\
         $V$ & $59023.53\pm0.20$ &$2.26\pm0.10$ & $2.38\pm0.05$ \\
    \hline\hline\\
    \end{tabular}
    \begin{justify}
    {\small
    \textbf{Notes.} For \textit{n} ($t_0$) (Column~4), the weighted average of $t_0=59023.3\pm0.1$ is taken. We also show the values when $t_0$ was fit as a free parameter for each band, \textit{n} ($t_{0,\mathrm{band}}$) (Column~3).}
    \end{justify}
    \label{tab:rise}
\end{table}

The SN\,2020nlb pre-discovery non-detection from ATLAS was $>19.7$\,mag on 59023.28\,MJD, and it was also not detected by ZTF on 59023.22\,MJD. Both of these non-detections are around the same epoch as our estimated time of first light, meaning the SN would be expected to be extremely faint, and these non-detections could even be prior to first light. The Itagaki Astronomical Observatory’s 0.35 m telescope in Okayama, Japan observed the field on MJD\,59024.57 and the SN was not detected down to a limiting magnitude of 18.5 \citep{2021ApJ...922...21S}. Our power-law fits would predict magnitudes of $B=19.3$ and $V=18.2$ at this time, so this does provide some tension with our power-law fits (see also discussion and figure in Appendix~\ref{a1} for more details). It is possible that the SN may have displayed something more akin to an approximately linear very early rise phase like has been suggested for\ SN\,2013dy and SN\,2014J, for example \citep{2013ApJ...778L..15Z,2014ApJ...783L..24Z}. If the SN did show such a phase, this  would likely mean our estimated first light date could be too early.

The calculated time of first light and the associated uncertainties ($59023.3\pm0.1$\,MJD) assume that the power law rise remains true even at the very earliest times. As previously noted, for some SNe~Ia with very early detections or very constraining upper limits, like SN\,2013dy and SN\,2014J, this has been shown not to be the case \citep{2013ApJ...778L..15Z,2014ApJ...783L..24Z}. For those objects, a broken power law has been found to better describe the early rise. In these cases, the initial two or three days after first light display an approximately linear rise in flux, before giving way to a power-law rise. While, as shown above, our observations are consistent with a single power-law rise for each filter, a short linear rise phase cannot be ruled out. If such a rise was present, the true first light date could be later that the $t_0=59023.3\pm0.1$\,MJD derived from our fits, and would in turn mean our first spectrum is even earlier than the 2.6\,d after first light that we estimate. For SN\,2020nlb, the uncertainty from the fitting of our data with the power law rise is small. The systematic uncertainty that will arise from the assumption of the nature of the rise (e.g.\ power law vs broken power law) will be considerably larger than this $\pm0.1$\,d uncertainty.

The relative differences in the fitted power-law rises from filter to filter shown in Table~\ref{tab:rise} do not seem to obviously correlate with the fits to the synthetic light curve of SN\,2011fe presented in Table~3 of \citet{2015MNRAS.446.3895F}. For example, in their fits, the \textit{B} band had the lowest power-law index and \textit{V} band the highest. For the time period where the rise of a SN~Ia is well described by power-law fits, if an object has a $(B-V)$ colour that is becoming bluer, the power-law index for the \textit{B} band must be higher than that of \textit{V}. For such SNe~Ia that have early rises well described by power laws, how the colours change during these epochs will be directly linked to the power-law indices. For example, the evolution of the $(B-V)$ colour at this epoch will be linked to the indices of the \textit{B} and \textit{V}-band flux early rises (i.e.\ $f\propto{t^{n(B)}}$ and $t^{n(V)}$ respectively) in the form: 

\begin{equation}
(B-V)=-2.5\times[n(B)-n(V)]\mathrm{log}(t)+\mathrm{constant}.
\end{equation}

Type Ia supernovae in the red sample of \citet{2018ApJ...864L..35S} get bluer over the first few days after first light, whereas their blue sample remain at an approximately constant colour during this phase. Therefore, based on colour evolution alone, it would be expected that $n(B)>n(V)$ for the red sample, and for their blue sample $n(B)\approx{n}(V)$ (although note that some of their blue sample do not show power-law rises). SN\,2011fe is in the red sample of \citet{2018ApJ...864L..35S}, and SN\,2020nlb also belongs to the same group (see Section~\ref{sec:cols}), meaning both should have $n(B)>n(V)$.

\begin{figure}[ht]
\centering
\includegraphics[width=\columnwidth]{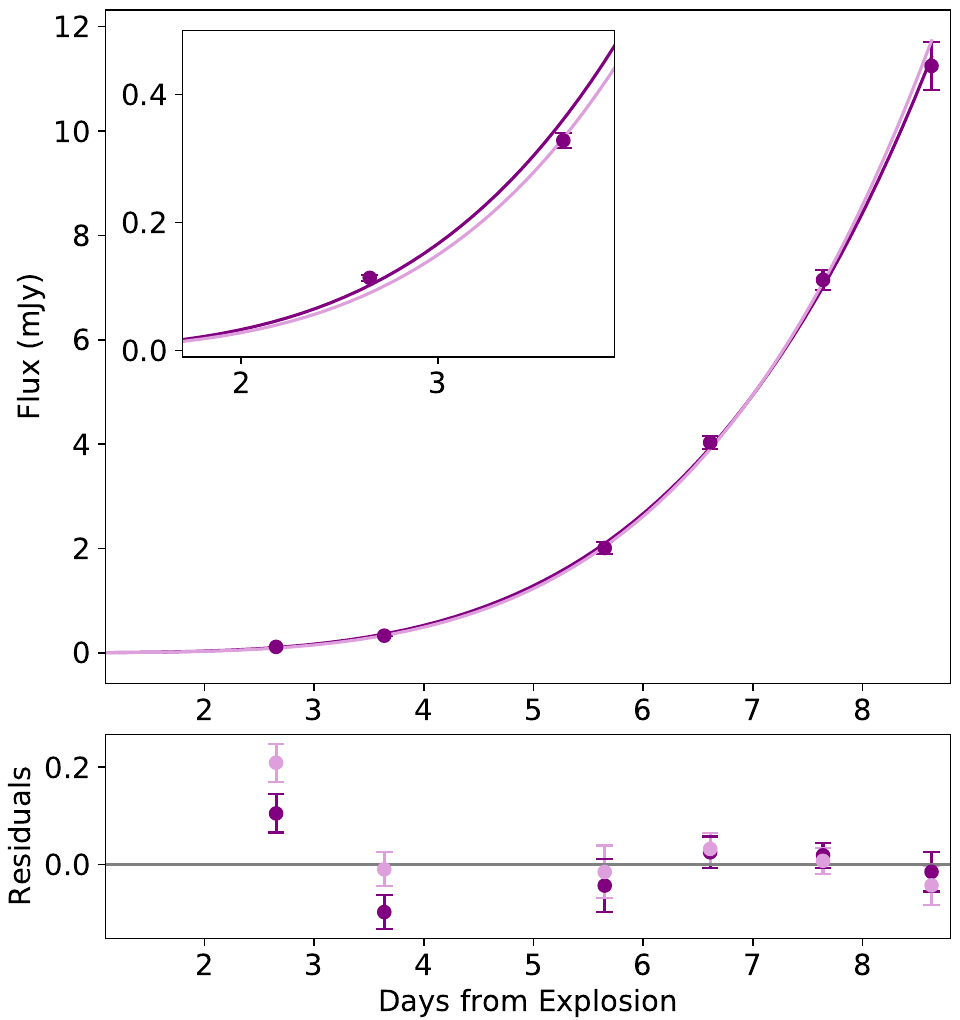}
\caption{Early $u'$-band light curve of SN\,2020nlb, showing the power-law fits with the first light date of 59023.3\,MJD. The dark purple fit and residuals show when all data used in Fig.~\ref{lce} is included, and the light purple shows the fit when the first epoch is excluded.}
\label{uearly}
\end{figure}

In recent years, several SNe~Ia have been found to show early excess flux that deviates from the traditional power-law rise. As discussed in the introduction, there are a number of possible mechanisms that could produce this, including interaction with a companion star, CSM, material from a WD merger, or excess surface $^{56}$Ni \citep{2010ApJ...708.1025K,2016ApJ...826...96P,2017MNRAS.472.2787N,2019ApJ...873...84P,2020arXiv200702101M}. Early photometric observations alone are not typically sufficient to distinguish between these, as many can potentially be tuned to match a variety of bumps. In these scenarios, the additional power source either comes from interaction or additional radioactive isotopes in the outer ejecta. There is no evidence of a substantial bump in the early light curve of SN\,2020nlb, and as can be seen in Fig.~\ref{lc}, power-law fits generally describe the early rise very well. The only possible exception to that is the $u'$ band. Examining the residuals shown in Fig.~\ref{lce}, we can see that the first $u'$-band data-point is marginally in excess of the best-fit power law, whereas the second data-point lies marginally below the best-fit, indicating the power-law fit from the derived first light time does not perfectly describe the $u'$-band early rise. We therefore re-fit the early $u'$-band data, but exclude the first epoch. This new fit is shown in Fig.~\ref{uearly} and compared to the fit that includes the first epoch. In the fit that excludes the first observation, the remaining data are well described by a power-law fit, and the first data point is then $\sim$5$\sigma$ in excess of that fit, with an excess flux-density of $23.8\pm4.5$\,$\mu$Jy.

\subsection{Colours}\label{sec:cols}

\begin{figure*}[ht]
\centering
\includegraphics[width=2\columnwidth]{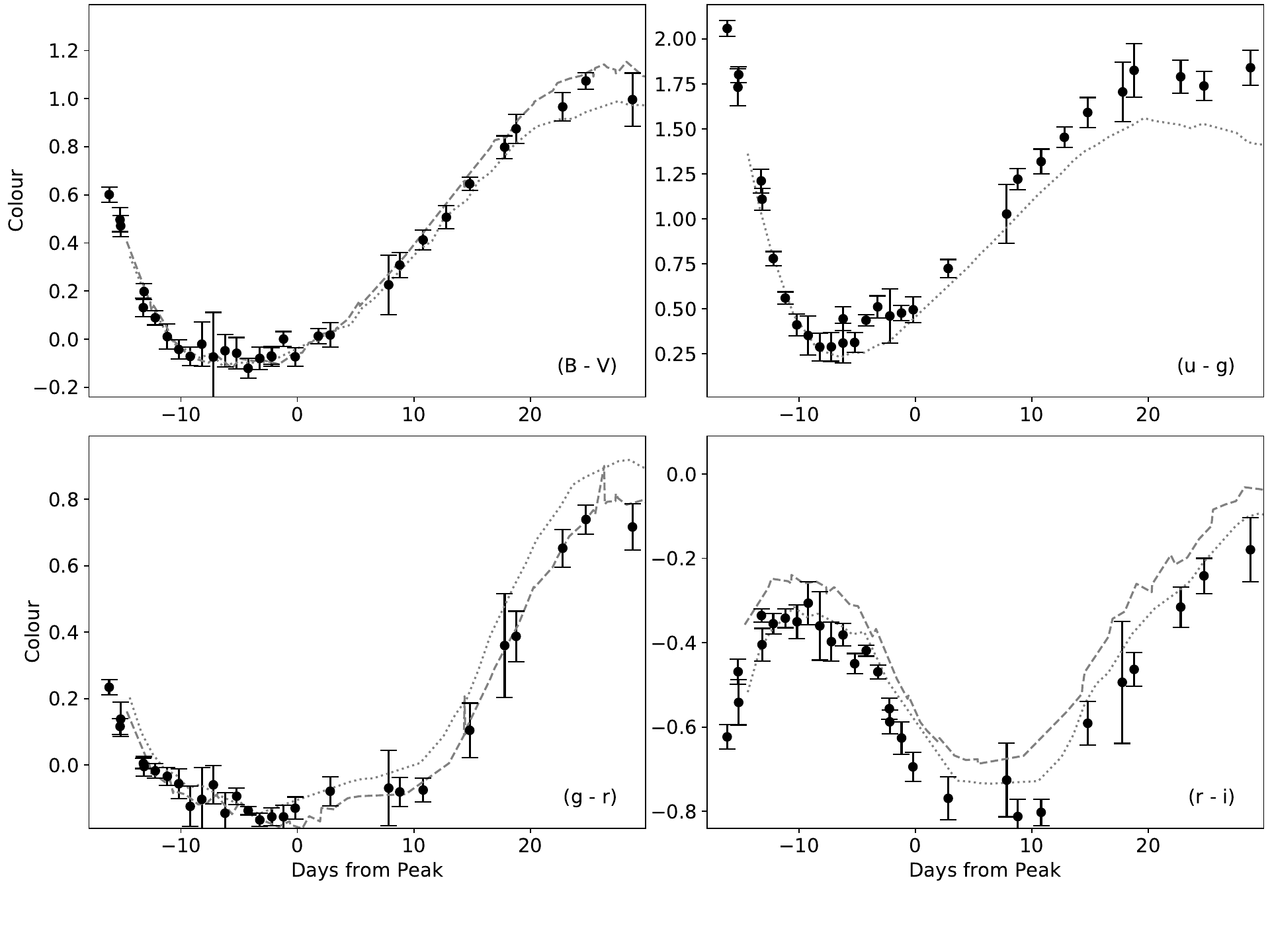}
\caption{Colour evolution of SN\,2020nlb (black points), compared to SN\,2015F photometry from \citet[][dashed grey line]{2017MNRAS.464.4476C} and \citet[][dotted grey line]{2018ApJ...869...56B}. The SN\,2020nlb colours are reddening corrected using the values discussed in the text. SN\,2015F is reddening corrected using the values from \citet{2017MNRAS.464.4476C}.}
\label{cols}
\end{figure*}

The early $(B-V)$ colour of SNe~Ia can vary by as much as $\sim$0.5\,mag, and they appear to fall into two groups, either showing red or blue early colours \citep{2018ApJ...864L..35S}. SN\,2020nlb belongs to the population showing red early $(B-V)$ colour. This is consistent with the findings of \citet{2018ApJ...864L..35S}, where the `blue' population showed only shallow silicon features in their spectra, unlike SN\,2020nlb. At the first photometry epoch, 2.6\,d after first light, we find $(B-V)=0.60\pm0.03$\,mag (correcting for foreground extinction of $E(B-V)=0.026$\,mag and taking $E(B-V)_{\mathrm{host}}=0.04$\,mag; see discussion in Section~\ref{sec:ext}), which then becomes steadily bluer until it reaches $(B-V)=-0.07\pm0.04$\,mag at 9.6\,d after first light, after which point it plateaus until around the time of maximum light. The colours of SN\,2020nlb are compared to those of SN\,2015F \citep{2017MNRAS.464.4476C,2018ApJ...869...56B} in Fig.~\ref{cols}.

\subsection{Extinction}\label{sec:ext}
The dust maps of \citet{2011ApJ...737..103S} indicate low Galactic reddening of $E(B-V)=0.026$\,mag towards the position of SN\,2020nlb. We adopt this value for the Galactic reddening, and assume $R_V=3.1$ \citep{1989ApJ...345..245C} throughout. Fitting our $u'BVg'r'i'$ photometry using `EBV$\_$model' in \texttt{SNooPy}, indicates a host galaxy reddening of $E(B-V)_{\mathrm{host}}=0.04\pm0.06$\,mag (the $\pm$0.06 is a systematic uncertainty which limits the accuracy with which \texttt{SNooPy} can determine reddening from SN~Ia colours, but obviously a negative extinction is unphysical). \texttt{SNooPy} derives the reddening value from the offset between the observed colours and those expected for that SN~Ia light curve. The spectra, which are presented in the next section, do not show strong Na~{\sc i}\,D absorption. There is a weak, but clear, Galactic component of Na~{\sc i}\,D, and the strength of the feature is roughly consistent (given the low resolution etc.) with that expected from the reddening of $E(B-V)=0.026$\,mag, using the relations from \citet{2012MNRAS.426.1465P}. We do not detect clear evidence of an M85 Na~{\sc i}\,D absorption feature. The low resolution of our spectra means that the D$_1$ and D$_2$ Na~{\sc i} lines are not resolved. This is further complicated by the low redshift of M85 ($z=0.0024$), which means that the Galactic and M85 Na~{\sc i}\,D features would not be completely resolved from each other. Nonetheless, estimating a conservative upper limit on the Na~{\sc i}\,D EW does allow us to confirm that $E(B-V)_{\mathrm{host}}<0.10$\,mag (i.e. the upper error bar of the $E(B-V)_{\mathrm{host}}$ derived in \texttt{SNooPy}) from the relations of \citet{2012MNRAS.426.1465P}. For the remainder of this work, we adopt a value of $E(B-V)_{\mathrm{host}}=0.04\pm0.06$\,mag. Although when we utilise a Monte Carlo technique to estimate physical parameters, we do not permit $E(B-V)_{\mathrm{host}}$ to be either negative or $>$\,0.1\,mag.

\subsection{Distance to M85} \label{dist}
{We use our light curve of SN\,2020nlb to estimate the distance to M85. Fitting the light-curve with \texttt{SNooPy} (using `EBV$\_$model') yields a distance modulus of $\mu_0=30.99\pm0.19$\,mag. This distance modulus corresponds to a distance of $15.8^{+1.4}_{-1.3}$\,Mpc. There are systematic uncertainties in the standardisation of SN~Ia light curves to measure the distance modulus, which as yet have been unable to be removed by additional parameterisation, and are typically of order $0.1-0.15$\,mag. This limits the accuracy to which one can determine the distance to an individual SN~Ia, even with a perfect dataset. M85 has previously hosted the SN~Ia 1960R \citep{1962MmSAI..33...77B}. This SN has been used in conjunction with other SNe~Ia to estimate the distance to the Virgo cluster \citep{1982ApJ...254....1A,1992ApJ...390L..45P}. However, the photographic observations of SN\,1960R itself have very poor early coverage \citep{1964AJ.....69..236B}, precluding the use of this SN to increase the accuracy of our SN~Ia-based distance determination for M85. Compared to our SN~Ia-derived value of $\mu_0=30.99\pm0.19$ (which includes systematic uncertainties), the PNLF gives a smaller distance modulus of $\mu_0=30.79\pm0.06$ \citep{1990ApJ...356..332J}, although this is consistent within the errors. At $\mu_0=31.26\pm0.05$ \citep{2007ApJ...655..144M}, SBF gives a larger distance than the SN~Ia value.}

\subsection{$^{56}$Ni mass} \label{sec:ni}
Following Arnett's rule \citep{1982ApJ...253..785A}, the mass of $^{56}$Ni synthesised in the explosion of a SN~Ia can be estimated from the peak of the bolometric light curve. By scaling our spectroscopic sequence to our photometric observations and then integrating the total flux, we can measure the total optical luminosity of the SN. The majority of the luminosity of a SN~Ia near maximum light is radiated in the optical. However, a significant fraction is emitted in the UV and NIR (for example, at \textit{B}-band maximum, SN\,2011fe emitted around 10\% and 6\% of its bolometric luminosity in the UV and NIR respectively; \citealp{2013A&A...554A..27P}). To account for this flux, we assume the same \textit{(i~--~NIR)} colours as for the similar SN~Ia SN\,2015F. We obtain the colours of that SN from \citet{2017MNRAS.464.4476C}. We then obtain \textit{JHK$_s$} magnitude estimates for SN\,2020nlb by applying these colours (after correcting for reddening; values from \citealp{2017MNRAS.464.4476C}) to our $i'$-band photometry of SN\,2020nlb. We then scale the NIR spectroscopic sequence of SN\,2011fe from \citet{2014MNRAS.439.1959M} to our estimated NIR photometry of SN\,2020nlb. For the UV contribution, we utilise the \textit{Swift UVM2} photometry of SN\,2020nlb published by \citet{2021ApJ...922...21S}. The \textit{UVM2} filter was chosen as it does not have a red tail to its transmission curve to the same extent that \textit{UVW1} and \textit{UVW2} have. The SED of a SN~Ia means that \textit{u}-band and even \textit{B}-band flux can significantly contribute to the measured flux in those filters, whereas such contamination is much lower for \textit{UVM2} (see Fig.~1 of \citealp{2016AJ....152..102B}). The SN\,2011fe UV spectra from \citet{2014MNRAS.439.1959M} were then scaled to \textit{UVM2} photometry for SN\,2020nlb. After adding these contributions, we integrated the flux between 1780\,\AA\ and 24980\,\AA, which we define as the bolometric luminosity.

\begin{figure*}
    \centering
    \includegraphics[width=\textwidth]{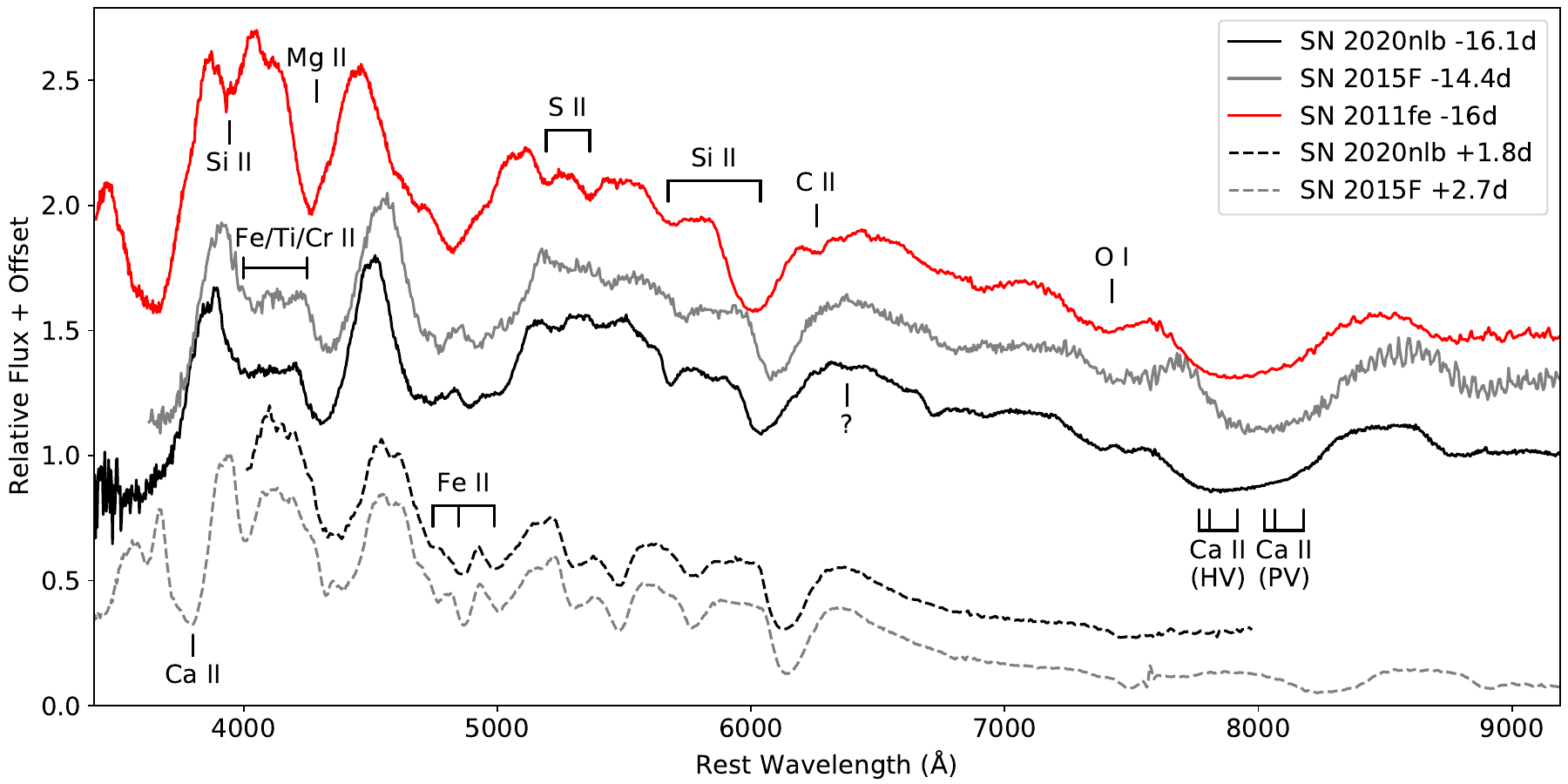}
    \caption{Spectroscopic comparison between early-time spectra of SN\,2020nlb, SN\,2015F \citep{2017MNRAS.464.4476C}, and SN\,2011fe \citep{2011Natur.480..344N}. Spectra of SN\,2020nlb and SN\,2015F taken near \textit{B}-band maximum are also shown. The positions of the high-velocity and photospheric-velocity components of the Ca~{\sc ii}\,NIR triplet are indicated, as fit from the first SN\,2020nlb spectrum (see text). The `?' label on the early SN\,2020nlb spectrum is an unidentified feature, which we determine is not likely to be C~{\sc ii} (see discussion in text).}
    \label{fig:15F}
\end{figure*}

The total mass of $^{56}$Ni synthesised in the explosion can be calculated from the peak bolometric luminosity, $L_{\mathrm{bol}}$, and the time between the explosion and the said peak, $t_{\mathrm{r}}$. We estimate the rise time (from explosion to bolometric peak) to be $17.2\pm0.3$\,d for SN\,2020nlb. From the equations in \citet{2005A&A...431..423S}, the $^{56}$Ni yield of the explosion can be calculated from:

\begin{equation}
M_{\mathrm{^{56}Ni}}=\frac{L_{\mathrm{bol}}}{\alpha[6.45\times{e}^{-t_{\mathrm{r}}/8.8}+1.45\times{e}^{-t_{\mathrm{r}}/111.3}]\times10^{43}}~\mathrm{M}_{\odot}.\label{nimass}
\end{equation}

\noindent Here $\alpha$ is the ratio between the radiated energy at maximum light and the instantaneous energy production from decays, where $\alpha=1$ is the special case in which the energy radiated is equal to the instantaneous energy production. For this work, we assume $\alpha=1$, except where directly stated otherwise.

Taking into account the uncertainty in the distance to M85, as derived in this work ($\mu_0=30.99\pm0.19$\,mag; see Section~\ref{dist}), we estimate a peak luminosity of SN\,2020nlb to be $(1.09^{+0.26}_{-0.17})\times10^{43}$\,erg\,s$^{-1}$. \citet{2021ApJ...922...21S} derive a peak bolometric luminosity of $\sim1\times10^{43}$\,erg\,s$^{-1}$. This is consistent with our value, although there are some differences in how the two values were obtained. For example, \citet{2021ApJ...922...21S} account for the NIR flux in a different way, use a larger distance to M85, and assume no host reddening.

After obtaining the peak bolometric luminosity and the bolometric rise time of SN\,2020nlb ($17.2\pm0.3$\,d), Equation~\ref{nimass} can be used to compute the $^{56}$Ni mass, which we derive to be $0.52^{+0.13}_{-0.08}$\,$M_{\odot}$. As these calculations use the distance derived from the SN light curve, they are fundamentally linked to the SN~Ia standardisation process, and to a substantial degree, will effectively be typical parameters for a supernova with this $\Delta{m}_{15}(B)$. If we instead use, for example, the independent distance modulus of $\mu_0=31.26\pm0.05$ \citep{2007ApJ...655..144M}  derived using the SBF method, this would increase the distance modulus by nearly 0.3\,mag. This would have a large effect on the derived $^{56}$Ni mass, increasing it to $M_{^{56}\mathrm{Ni}}=0.65^{+0.11}_{-0.04}$\,$M_{\odot}$. As previously noted, the calculated $^{56}$Ni mass of $0.52^{+0.13}_{-0.08}$\,$M_{\odot}$ assumes that $\alpha=1$. If for example, we instead take $\alpha=1.2$ \citep{1992ApJ...392...35B}, then the $^{56}$Ni mass would be $0.44^{+0.11}_{-0.07}$\,$M_{\odot}$. Finally, we can give a general $M_{^{56}\mathrm{Ni}}$ value which includes an uncertainty where $\alpha$ can be anywhere between 1 and 1.2, which results in $0.48^{+0.12}_{-0.08}$\,$M_{\odot}$.

{The maximum light spectra of SN\,2020nlb have quite a strong Si~{\sc ii} 5972\,\AA\ line, indicating a temperature on the lower side of the normal SN~Ia distribution, as would be expected for a SN~Ia with a lower luminosity than an average normal SN~Ia. This is not very compatible with a peak of $M_B\sim-19.4$\,mag and $^{56}$Ni mass in excess of 0.6\,$M_{\odot}$ that would be implied using the SBF distance under the assumption that the parameter  $\alpha=1$. Therefore, as well as the light-curve fitting, the spectra of SN\,2020nlb, combined with the estimated bolometric flux at maximum light, favour a distance to M85 lower than the SBF-derived value. If there is no host galaxy extinction, only the Galactic $E(B-V)=0.026$\,mag, the SBF distance modulus of $\mu_0=31.26\pm0.05$ would imply $M_B=-19.20$ and $^{56}$Ni mass of $0.61\pm0.03$\,$M_{\odot}$. This still seems on the high side given the spectroscopic evolution of SN\,2020nlb, but given the implied $^{56}$Ni mass would fall to $0.51\pm0.02$\,$M_{\odot}$ if $\alpha=1.2$, it does then become reconcilable. Therefore we cannot definitively say that SN\,2020nlb is incompatible with the SBF-derived distance modulus.}

\section{Spectroscopy} \label{sec:spec}

The spectroscopic evolution of SN\,2020nlb around peak brightness is typical of a normal SN~Ia. However, our early high S/N spectroscopic follow-up probes epochs rarely observed in SNe~Ia. In the subsequent analysis and discussion of the photospheric-phase spectra, we therefore focus primarily on these earliest times.

\begin{figure*}[ht!]
    \centering
    \includegraphics[width=2\columnwidth]{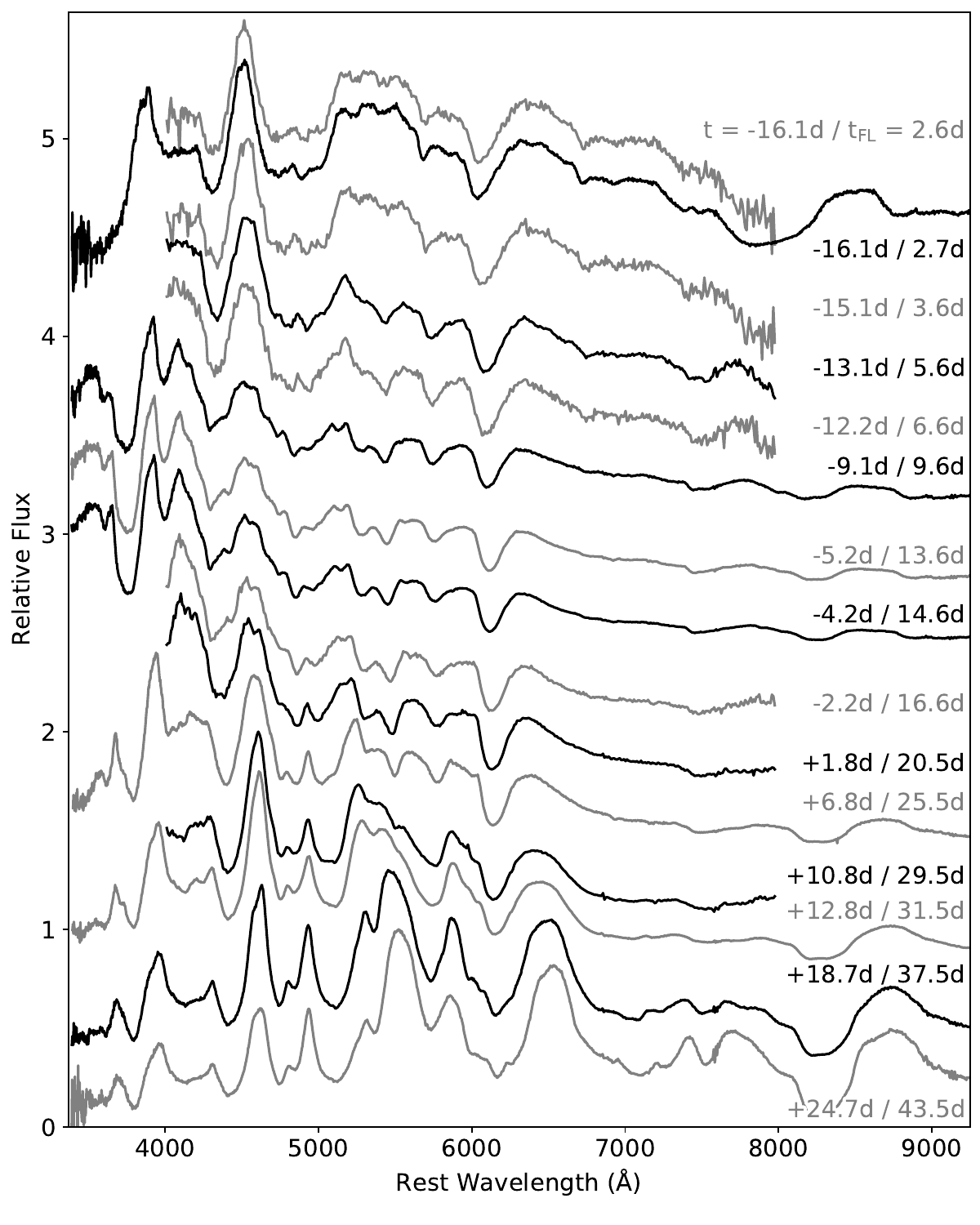}
    \caption{Our ALFOSC and SPRAT spectra of SN\,2020nlb, with the phase with respect to the \textit{B}-band maximum (\textit{t}) and calculated time of first light ($t_{\mathrm{FL}}$) indicated. The alternating black and grey colours are simply for a visual aid. No extinction correction has been applied to these spectra.}
    \label{fig:spec}
\end{figure*}

\subsection{Spectroscopic evolution}
Our first two spectra, by the LT and NOT, were taken just 0.05\,d apart (rounded to 2.6 and 2.7\,d after first light respectively). The slightly later NOT spectrum has both superior S/N and broader wavelength coverage. We therefore focus our analysis of the earliest times on this NOT spectrum. The spectrum was taken when the SN was still 4.9\,mag fainter than \textit{B}-band maximum. The spectrum shows low excitation lines, including strong Fe~{\sc ii} absorption, as well as Ti~{\sc ii} absorption features, which are more typically associated with 1991bg-like or transitional SNe~Ia. It is similar to the earliest spectrum of SN\,2015F \citep{2015ATel.7209....1F,2017MNRAS.464.4476C}.  The early-time spectrum of SN\,2020nlb is very different to the early spectra of SN\,2011fe \citep{2011Natur.480..344N}. These comparisons are shown in Fig.~\ref{fig:15F}. The cool spectrum of SN\,2020nlb persists for a few days, meaning an uncertainty in the true explosion time would be unable to explain the differences between SN\,2020nlb and SN\,2011fe. At 2.7\,d after the estimated time of first light, our first NOT spectrum of SN\,2020nlb is among the earliest high-quality spectra taken of a SN~Ia. Our initial spectroscopic sequence, which is shown in Fig.~\ref{fig:spec}, begins $16.1$\,d prior to \textit{B}-band maximum and ends $24.7$\,d after that peak.

\begin{figure}
    \centering
    \includegraphics[width=\columnwidth]{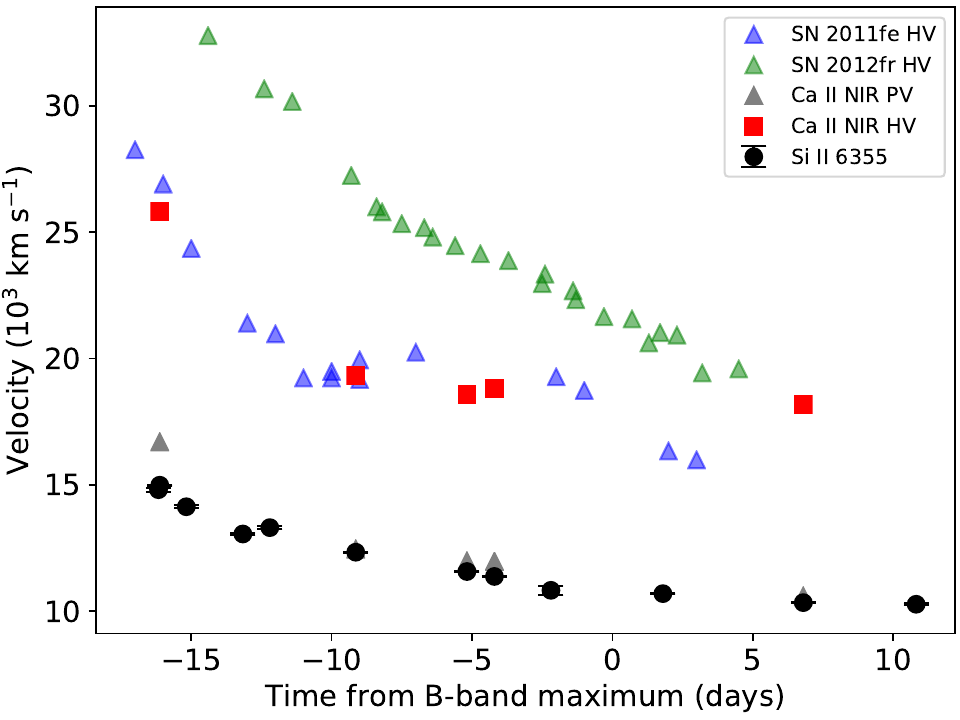}
    \caption{SN\,2020nlb velocity evolution of the Si~{\sc ii} 6355\,\AA\ line, and the HVF and PVF of the Ca~{\sc ii}~NIR triplet. Comparisons to the Ca~{\sc ii}~NIR HVF of SN\,2011fe and SN\,2012fr are also shown (data from \citealp{2015MNRAS.451.1973S}).}
    \label{fig:vel}
\end{figure}

In the first spectrum, the Ca~{\sc ii}~NIR triplet absorption feature extends to very high velocities and merges with the absorption features corresponding to O~{\sc i} 7773\,\AA\ and Mg~{\sc ii} 7889\,\AA. The initial Si~{\sc ii} 6355\,\AA\ velocity is measured at $14900\pm100$\,km\,s$^{-1}$ (this velocity is derived from the minimum of the absorption feature). At these times there is lower flux in the region between the  5972 and  6355\,\AA\ Si~{\sc ii} lines. It is possible that this is caused by high-velocity Si or another species. Given the strength of the Fe~{\sc ii} lines in other parts of the spectrum, Fe~{\sc ii} likely significantly influences this region (see discussion in Section~\ref{sec:synapps}). Our spectroscopic sequence begins very early, so a Ca~{\sc ii}~NIR triplet high-velocity feature (HVF) would be expected to be present, whereas prominent Si~{\sc ii} HVFs are less common \citep{2005ApJ...623.1011B,2005ApJ...623L..37M}. We measure the velocity of O~{\sc i} at the same epoch to be $14900\pm100$\,km\,s$^{-1}$ and Mg~{\sc ii} to be $15200\pm100$\,km\,s$^{-1}$.

The Si~{\sc ii} velocity evolution is shown in Fig~\ref{fig:vel}. In order to map the Ca~{\sc ii}~NIR velocity evolution, and the HVF in particular, we fit Gaussian profiles to the line. The HVF and photospheric-velocity feature (PVF) are fit as two separate features. Each of the two features are fit as three Gaussian profiles, corresponding to Ca~{\sc ii} 8498, 8542 and 8662\,\AA, with both the width and velocities of the three lines fixed (in velocity space) with respect to each other. In the fitting process, the relative strengths of the individual 8498, 8542 and 8662\,\AA\ lines within each component are permitted to vary between that expected from the log~\textit{gf} values and unity. The velocity evolution of both the Ca~{\sc ii}~NIR HVF and PVF are shown in Fig.~\ref{fig:vel}. The early $t=-16.1$\,d spectrum (2.7\,d after first light) yields a best-fit velocity of 25800\,km\,s$^{-1}$ for the HVF. The blue edge of the Ca~{\sc ii}~NIR feature is at a considerably higher velocity than this. It is not possible to accurately measure this, due to that end of the feature being blended with Mg~{\sc ii} and O~{\sc i}, however we find that Ca~{\sc ii} remains the strongest component of this blended feature out to a velocity of 33000\,km\,s$^{-1}$. The best-fit velocity of the Ca~{\sc ii}~NIR HVF falls to 19300\,km\,s$^{-1}$ by $t=-9.1$\,d, when the second ALFOSC spectrum was taken. We can see some early evolution of the Ca~{\sc ii}~NIR HVF by examining the red end of the SPRAT spectra taken between these two ALFOSC spectra (see Fig.~\ref{fig:spec}). Due to the more limited wavelength coverage, the full Gaussian fitting described above is not possible. Nonetheless, it can be seen that by $t=-13.1$\,d, the velocity of the blue edge of the feature has already fallen substantially, as in the $t=-13.1$\,d SPRAT spectrum, it is now clearly separate from the neighbouring O~{\sc i} 7773\,\AA\ and Mg~{\sc ii} 7889\,\AA\ that it had been merged with at the earliest epochs. We can see from Fig.~\ref{fig:vel} that the velocity evolution of the Ca~{\sc ii}\,NIR HVF in SN\,2020nlb is consistent with that of SN\,2011fe, showing a very rapid fall in velocity at early phases, before effectively plateauing from $\sim$10\,d prior to maximum light. This is in contrast to the shallower and more prolonged Ca~{\sc ii}\,NIR HVF decline in SN\,2012fr \citep{2015MNRAS.451.1973S}. The Ca~{\sc ii}~NIR HVF also weakens quite rapidly with time, and by the spectrum taken 4.2\,d prior to maximum light, the HVF is weak in comparison to the photospheric component of the line. Despite the Ca~{\sc ii}\,NIR HVF being strong at early times, SN\,2020nlb does not show unambiguous signs of a Si~{\sc ii} HVF (but it may be present, see discussion in Section~\ref{sec:synapps}). The same is true for SN\,2011fe and SN\,2015F, unlike the more luminous SN\,2012fr, which shows strong HVF of both Si~{\sc ii} and Ca~{\sc ii} \citep{2013ApJ...770...29C}.

We find a Si~{\sc ii} velocity gradient of $50\pm5$\,km\,s$^{-1}$\,day$^{-1}$ and a maximum-light Si~{\sc ii} velocity of $10750\pm50$\,km\,s$^{-1}$. The SN is in the low-velocity gradient classification of \citet{2005ApJ...623.1011B}. In the spectrum taken shortly after maximum light, we measure the equivalent widths of the Si~{\sc ii} 5972 and 6355\,\AA\ lines to be 24 and 116\,\AA\ respectively. These measurements place SN\,2020nlb near the border of the `core-normal' and `broad-lined' SNe~Ia in the classification scheme of \citet{2006PASP..118..560B} and it is in the `normal' classification of \citet[][see also \citealp{2012AJ....143..126B}]{2009ApJ...699L.139W}.

Our spectroscopic sequence of SN\,2020nlb shows no clear evidence of the C~{\sc ii} 6580\,\AA\ line. In the first ALFOSC spectrum, there does appear to be a weak absorption feature atop of the peak reward of the Si~{\sc ii} 6355\,\AA\ absorption feature. However, if this was C~{\sc ii}, it would imply a velocity of $\sim$9000\,km\,s$^{-1}$, extremely low with respect to the velocities seen in the main absorption spectrum (e.g.\ Si~{\sc ii}, Fe~{\sc ii}; $\sim$15000\,km\,s$^{-1}$) at this time. This would be lower even than the Si~{\sc ii} velocity at maximum light, and difficult to reconcile. There does however appear to be an absorption line an the approximate position that would be expected for C~{\sc ii} 7234\,\AA. This must presumably be another species if the 6580\,\AA\ line is absent, unless it is the case that the 6580\,\AA\ line is sufficiently hidden by other features such as Si~{\sc ii} and Fe~{\sc ii}.

\citet{2019ApJ...871..250H} explored the effect that various carbon mass fractions in the velocity range of $7850-16000$\,km\,s$^{-1}$ has on the pre-maximum spectra of SNe~Ia. For the inner portion of this velocity range, even a fairly small C mass fraction can create a C~{\sc ii} 6580 feature that is clearly visible. However for the higher part of this velocity range ($13500-16000$\,km\,s$^{-1}$), the lack of a clear C~{\sc ii} feature is less constraining. Given the complication of the feature being near the Si~{\sc ii} line, it does not seem implausible that a C mass fraction of up to 0.01 could be present in this velocity range of the ejecta without a clear indication in the optical spectra. It is worth noting that in a cool early spectrum with strong absorption from singly ionised metal lines, as is the case for SN\,2020nlb, there is also potential ambiguity near C~{\sc ii}~ 6580\,\AA. For example, Fe~{\sc ii} and Ti~{\sc ii} can also produce absorption in the same vicinity. The presence and strength of the C~{\sc ii} lines will also depend on the ionisation of the carbon, which will be lower for lower-luminosity SNe~Ia \citep{2014MNRAS.441..532D}.

\subsection{Spectral fitting}\label{sec:synapps}

\begin{figure*}
    \centering
    \includegraphics[width=2\columnwidth]{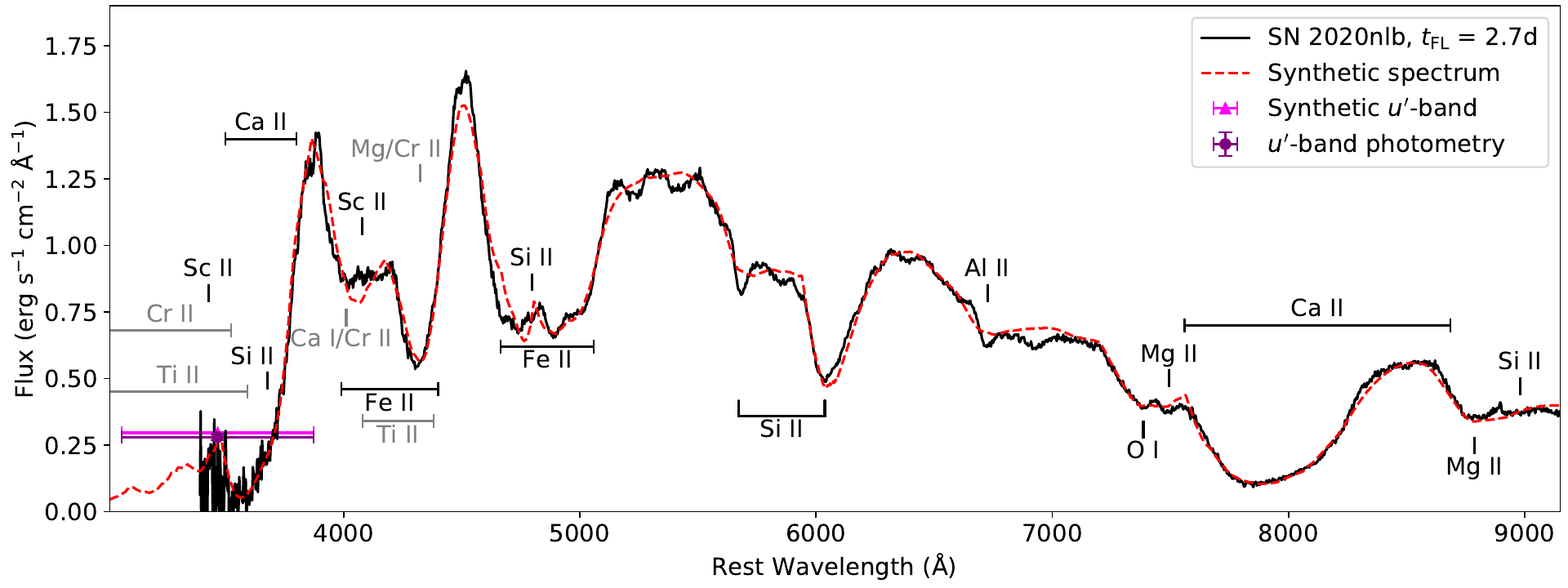}
    \caption{Synthetic \texttt{SYNAPPS} fit to our first NOT spectrum of SN\,2020nlb. We indicate regions with strong features of various species, as implied by the spectral fitting. Synthetic $u'$-band photometry was performed on the \texttt{SYNAPPS} spectrum, which is shown on the plot, as is the actual $u'$-band photometry for this epoch.}
    \label{fig:syn}
\end{figure*}

We use \texttt{SYNAPPS} SN spectral fitting code \citep{2011PASP..123..237T} to analyse our spectra, focusing in particular on the earliest NOT spectrum. Our fit is shown in Fig.~\ref{fig:syn}, with line identifications for many of the strongest features in the synthetic spectrum also shown. \texttt{SYNAPPS} uses simple assumptions of the ejecta structure and line formation to produce synthetic spectra. The contributions of different ions can then be fit iteratively to the data for a given SN. Such codes do not yield the quantitative abundance estimates that are possible with radiative transfer codes. However, they are useful for detecting the presence of different ions and from that a qualitative understanding of the overall ionisation and excitation of the SN ejecta.

As mentioned earlier, there is a lack of flux between the Si~{\sc ii} 5972 and 6355\,\AA\ lines in the first spectrum. There is possibility that this is due to a Si~{\sc ii} 6355\,\AA\ HVF. This possible HVF of Si~{\sc ii} 6355\,\AA\ would be at $\sim$24000\,km\,s$^{-1}$. This would be around 2000\,km\,s$^{-1}$ offset from the Ca~{\sc ii} HVF, but substantial differences between the Si and Ca HVF velocities would not be particularly unusual (see e.g.\ \citealp{2015MNRAS.451.1973S}). We also note that the PVF of the two ions are offset. This is not a secure identification, as the feature could be from low-excitation species that quickly disappears as the spectrum becomes hotter on the rise to peak. Given the strength of the Fe~{\sc ii} features in this early spectrum, Fe~{\sc ii} 6148\,\AA\ likely contributes significantly to this line, although our \texttt{SYNAPPS} fitting of the whole spectrum indicates that it may not be strong enough to be the sole species responsible for the line. The \texttt{SYNAPPS} fits indicate O~{\sc i} 6157\,\AA\ should not make a substantial contribution. One possibility could be Ca~{\sc i} Multiplet 3 (6102, 6122, 6162\,\AA), which could produce a feature here, and has been identified in spectra of SN\,1991bg-like SNe~Ia near maximum light \citep{2004ApJ...613.1120G,2011PASP..123..765D}. At this very early epoch, the spectrum of SN\,2020nlb actually shows many similarities to a SN\,1991bg-like SN~Ia near maximum light (with the obvious exception of the much higher velocities), and the ionisation balance of the ejecta in the line-forming region is probably not so different. However, it is difficult to make an identification with any confidence because of the fact that the other strong Ca~{\sc i} optical lines are mainly at the blue end of the spectrum, where there will be degeneracies with other species in the fitting due to the heavy line blanketing from multiple overlapping species at this epoch. We stress that \texttt{SYNAPPS} simply shows that Ca~{\sc i} itself does not produce any lines that would be inconsistent with the early spectrum, it is not able to assess the plausibility of different ions being present in the ejecta and their relative strengths. We also note that due to the heavy line blanketing at the blue end of the spectrum from a multitude of overlapping lines, there are likely to be substantial degeneracies between the strengths of different ions. Any ion that does not have a strong feature red-wards of 4500\,\AA\ would be especially susceptible to this.

We briefly discussed earlier the feature that may correspond to C~{\sc ii} 7234\,\AA, but for that identification to be correct, it would have to be the case that the 6580\,\AA\ line is sufficiently hidden by other features such as Si~{\sc ii} and Fe~{\sc ii}. At least within our \texttt{SYNAPPS} fitting, we are unable to successfully do this. Whenever the C~{\sc ii} 7234\,\AA\ line is sufficiently strong within the fitting to match the actual spectrum, a clear C~{\sc ii} 6580\,\AA\ line is also produced, which is not visible in the spectrum.

The S~{\sc ii} `W' feature (5460 and 5640\,\AA) is not obvious at the earliest times. There are two weak absorption troughs in the vicinity, with the approximate spacing expected for the two components of the S~{\sc ii} `W', but if these are indeed S~{\sc ii}, it would be at a velocity $\sim$12500\,km\,s$^{-1}$, substantially lower than the other isolated lines at the same epoch, Si~{\sc ii}, O~{\sc i} and Mg~{\sc ii}, which are all at $\sim$15000\,km\,s$^{-1}$, and the PVF of Ca~{\sc ii} at this time is $\sim$16700\,km\,s$^{-1}$.

\begin{figure*}[ht!]
    \centering
    \includegraphics[width=2\columnwidth]{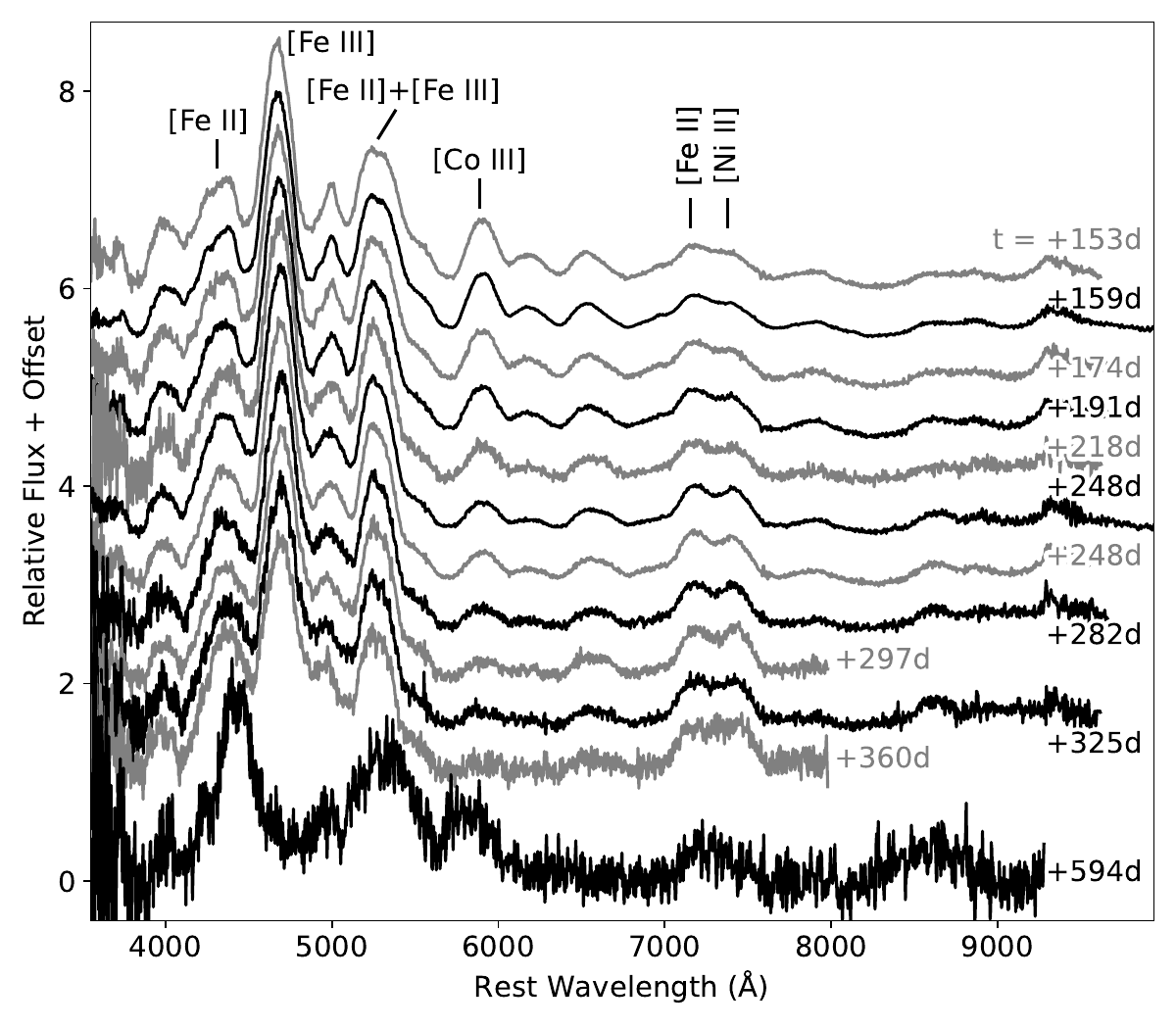}
    \caption{Our ALFOSC, LRS, and FORS2 nebular spectra of SN\,2020nlb, with the phase with respect to the \textit{B}-band maximum indicated.}
    \label{fig:neb}
\end{figure*}

The fitted ions need to suppress the flux in the $3400-3500$\,\AA\ range, corresponding to the region blue-wards of the Ca~{\sc ii} H~\&~K HVF. This requirement can be seen by comparing the early spectra of SN\,2020nlb to SN\,2011fe in Fig.~\ref{fig:15F}, where the Ca~{\sc ii}\,NIR features are broadly similar in the two SNe, yet the region blue-wards of Ca~{\sc ii}\,H~\&~K rises in flux for SN\,2011fe, but is still almost completely suppressed in SN\,2020nlb. In addition, the flux in the remainder of the $u'$ band, $3100-3400$\,\AA, must also be fairly low if it is to match the observed $u'$-band photometry at the epoch. The $u'$-band photometry is shown in Fig.~\ref{fig:syn}, and the synthetic spectrum is extrapolated to the blue edge of the $u'$ band (using the ion fits from the rest of the spectrum). We can see that the synthetic photometry of our \texttt{SYNAPPS} fit to the first spectrum is consistent with the actual $u'$-band photometry at the same epoch. In addition to Ca~{\sc ii} and Si~{\sc ii}, the blue end of the spectrum is dominated by multiple overlapping lines of IGEs. Our \texttt{SYNAPPS} fitting shows that the spectrum can be replicated by including Fe~{\sc ii}, Ti~{\sc ii}, Cr~{\sc ii}, Sc~{\sc ii} and Ni~{\sc ii}. As previously stated, it is possible to use Ca~{\sc i} to account for the feature between the 5972 and 6355\,\AA\ Si~{\sc ii} lines, but that identification is speculative. We observe an absorption line at 6720\,\AA, which was identified as either Al~{\sc ii} or high-velocity C~{\sc ii} in SN\,2015F by \citet{2017MNRAS.464.4476C}. Some models predict Al~{\sc ii} features in early SN~Ia spectra \citep{2015MNRAS.448.2766B}, so its presence here at early epochs would not be a surprise, and we use this in our fits.

From the ions included in the fitting, there is no clear indication as to the identification of the relatively weak line at $\sim$6390\,\AA\ (the line labelled as `?' in Fig.~\ref{fig:15F}). It is only unambiguously detected in the first spectrum, and has clearly disappeared three days later, in the $t=-13.1$\,d spectrum. It therefore must presumably be either from a species only present in the outer ejecta, or from a low-excitation line. 

There is substantial evolution of the broad 4800\,\AA\ complex between $t=-13.1$\,d and $t=-9.1$\,d, where it transitions from being dominated by metal lines, particularly Fe~{\sc ii}, to being more strongly influenced by Si~{\sc ii} and S~{\sc ii}. This will primarily be due to a move towards higher temperatures. The observable ejecta at this earliest epoch will be largely uncontaminated by iron synthesised in the explosion, as the IGEs will be primarily located at deeper layers, and only a small faction ($<$1\%) of the $^{56}$Ni will have decayed to iron in any case. In addition to contributing more to the 4800 feature, the S~{\sc ii} `W' feature also becomes much clearer between these epochs. The spectra between $t=-9.1$\,d and $t=-4.2$\,d are broadly similar. Changes between these epochs include a clearer Si~{\sc iii} 4559\,\AA\ feature. Comparing the $t=-16.1$\,d and $t=+6.8$\,d ALFOSC spectra of SN\,2020nlb in Fig.~\ref{fig:spec}, the dominant features are not dissimilar, with the obvious exception of the much higher velocities in the earlier spectrum. At $t=+6.8$\,d and onwards, the spectrum becomes increasingly dominated by IGE lines, as is expected for a SN~Ia.

\subsection{Nebular spectroscopy}
We obtained a sequence of spectra as SN\,2020nlb entered the nebular phase, through to about one year after maximum. We then obtained a very late nebular spectrum of the SN at 594\,d after peak. Our nebular spectra are shown in Fig.~\ref{fig:neb}. Very few SNe~Ia have nebular spectroscopy at such late epochs. One SN~Ia that does have such late observations is SN\,2011fe. In addition, a few SNe~Ia have been observed spectroscopically at very late phases, but the flux at those times has been dominated by a light echo of the explosion, rather that the ejecta itself (e.g.~SN\,1991T and SN\,1998bu; \citealp{1994ApJ...434L..19S,2001ApJ...549L.215C}).

\begin{figure}[ht]
\centering
\includegraphics[width=\columnwidth]{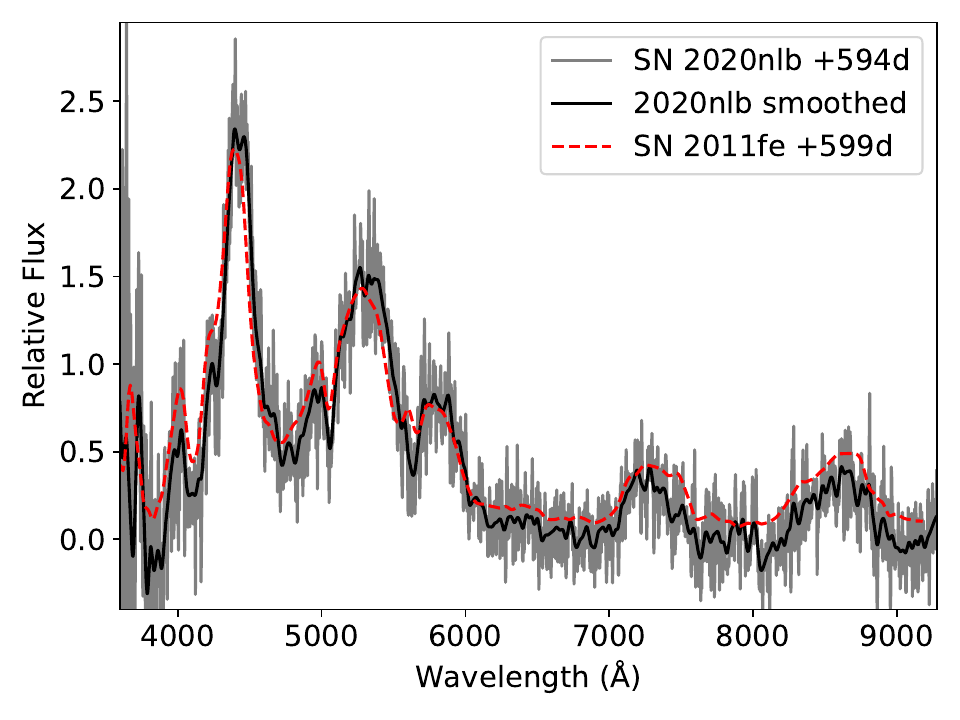}
\caption{Our final nebular spectrum of SN\,2020nlb, taken 594\,d after maximum light, compared to SN\,2011fe \citep{2022ApJ...926L..25T} at a similar phase. The spectra have been normalised to the flux region 4200--6000\,\AA, and reddening corrected using the values in Table~\ref{tab:nebref}.}
\label{fig:lateneb}
\end{figure}

Our very late spectrum of SN\,2020nlb is compared to SN\,2011fe at a similar phase in Fig.~\ref{fig:lateneb}, which shows the two spectra to be very similar at this epoch. The [Fe~{\sc iii}] lines, which were previously very prominent, have by this point disappeared, as the iron in the ejecta has shifted to lower ionisation. This shift in ionisation has been suggested as being due to clumping in the ejecta (\citealp{2020MNRAS.494.2809M,2022ApJ...926L..25T}; see also \citealp{2020MNRAS.494.2221W}). As well as the disappearance of [Fe~{\sc iii}], there are some other changes to the spectrum, including a substantial red-ward shift of the line previously around 4300\,\AA, with the line now being centred around 4420\,\AA, suggesting another species is contributing to the flux. This shift was also seen in SN\,2011fe and has been suggested as being due to [Fe~{\sc i}] emission. \citet{2015ApJ...814L...2F} suggested the blue Fe complex in SN\,2011fe was primarily [Fe~{\sc i}] emission by 1000\,d after peak, with [Fe~{\sc ii}] only making a relatively small contribution. 

\begin{figure}[ht]
\centering
\includegraphics[width=\columnwidth]{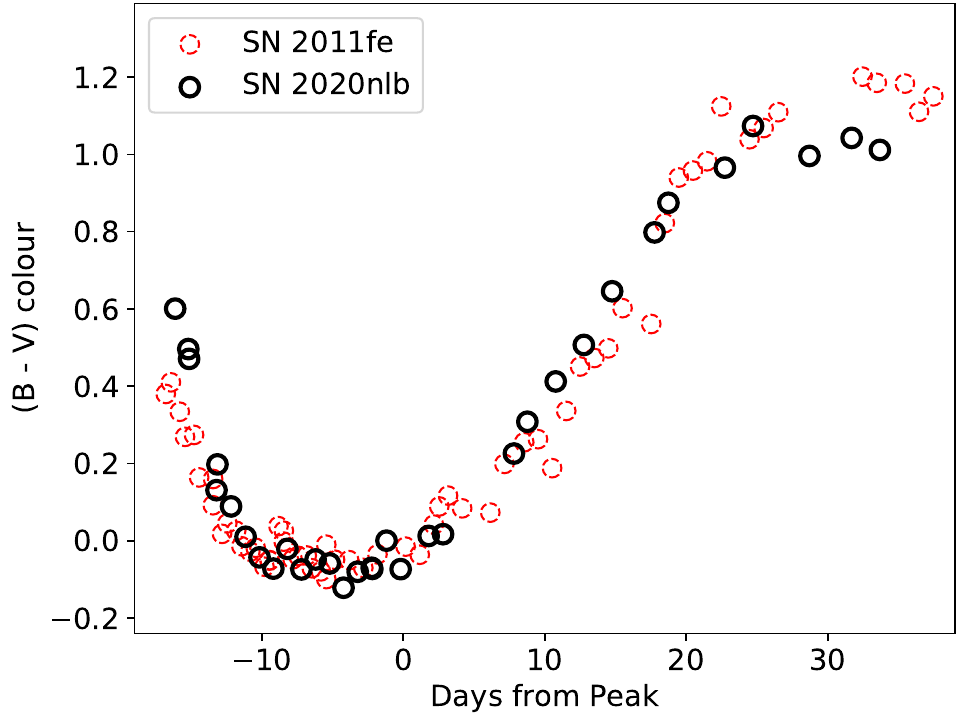}
\caption{$(B-V)$ colour of SN\,2020nlb compared to SN\,2011fe \citep{2016ApJ...820...67Z}, the two SN~Ia with very late nebular spectroscopy. The colours have been reddening corrected.}
\label{latelc}
\end{figure}

The $(B-V)$ colour evolution of SN\,2011fe and SN\,2020nlb are compared in Fig.~\ref{latelc}. To minimise systematics, we re-calculate maximum light  parameters from the published photometry (data from \citealp{2016ApJ...820...67Z}) in exactly the same manner as we did for SN\,2020nlb, yielding $\Delta{m}_{15}(B)=1.14\pm0.02$\,mag for SN\,2011fe. So with $\Delta{m}_{15}(B)=1.28\pm0.02$\,mag, the decline shown by SN\,2020nlb is faster than that of SN\,2011fe. The most striking difference in the colours is the very early time observations, where SN\,2020nlb is substantially redder than SN\,2011fe. This is reflected in the spectral features, where SN\,2020nlb shows a cooler spectrum and the blue end is dominated by absorption from singly ionised metals, whereas SN\,2011fe shows more prominent IME features even at the earliest times.

\begin{figure*}[ht]
\centering
\includegraphics[width=2\columnwidth]{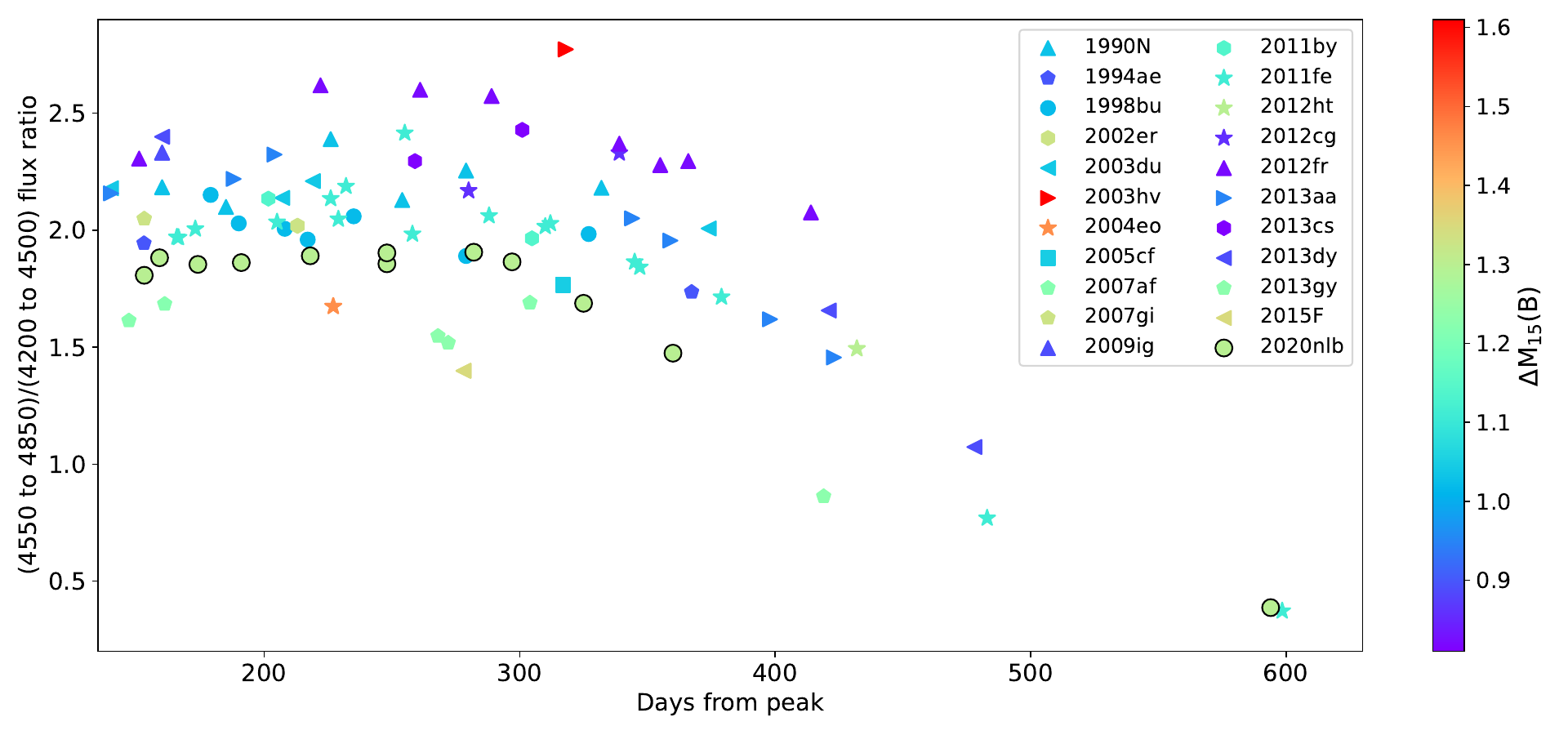}
\caption{Flux ratio of $(4550-4850)/(4200-4500)$\,\AA\ for various SNe~Ia observed in the nebular phase. Around one year after maximum, the numerator of the ratio is typically dominated by [Fe~{\sc iii}] flux and the denominator is dominated by [Fe~{\sc ii}] flux. All fluxes have been redshift and reddening corrected. For reddening values and references, see Table~\ref{tab:nebref}.}
\label{fig:ratio}
\end{figure*}

In Fig.~\ref{fig:ratio}, we show the flux ratio between the regions $4550-4850$\,\AA\ and $4200-4500$\,\AA, where the first listed region is dominated by [Fe~{\sc iii}] and the second range is dominated by [Fe~{\sc ii}] emission typical of SN~Ia nebular spectra at around 200--400\,d after maximum light. Therefore one would expect a higher $(4550-4850)/(4200-4500)$ flux ratio when the ejecta were at higher ionisation. Only spectroscopically normal SNe~Ia are included in the plot. For these objects, $\Delta{m}_{15}(B)$ is a relatively good proxy for the $^{56}$Ni mass, which is not necessarily the case for some over-luminous peculiar SNe~Ia. Additionally, some sub-luminous peculiar SNe~Ia show much narrower nebular lines than normal SNe~Ia (e.g.~SN\,1991bg; \citealp{1996MNRAS.283....1T}). Fig.~\ref{fig:ratio} shows this ratio with respect to the epoch after peak brightness for various spectroscopically normal SNe~Ia (see Table~\ref{tab:nebref} for parameters of each SN). It is difficult to get accurate uncertainties on the measured ratios, as the main source of error will be differences in the calibration, and the spectra come from many different sources and instruments. By looking at the SNe with multiple measurements in the 200--300 day range, the scatter in these suggests a 5\% error would be a reasonable approximation for the typical error on individual measurements of the flux ratio. From the figure, some trends are apparent:

\begin{enumerate}
    \item Generally, slower declining (and thus more luminous) SNe~Ia appear to generally have a higher $(4550-4850)$\AA$/(4200-4500)$\AA\ flux ratio.
    \item This is not the case for SN\,2003hv, which at $\Delta{m}_{15}(B)=1.61$\,mag is the fastest fading SN in the plot, yet  it exhibited the highest ratio of all.
    \item The best observed SNe with multiple spectra after 300\,d (SN\,2011fe, SN\,2012fr, SN\,2013aa, SN\,2020nlb) all show a decline in the flux ratio around the 300 to 400\,d period.
\end{enumerate}

There is a general trend where SNe with lower $\Delta{m}_{15}(B)$ values (i.e.~higher $^{56}$Ni yields) have higher ionisation in the nebular phase up to one year after peak. A big outlier in this trend is SN\,2003hv, with $\Delta{m}_{15}(B)=1.61$\,mag, but very high [Fe~{\sc iii}]/[Fe~{\sc ii}] ratio. The high ionisation seen in this particular SN in the nebular phase has been noted in the literature and it has been suggested as being due to lower density and possibly originating from a sub-$M_{\mathrm{Ch}}$ explosion \citep{2011MNRAS.416..881M}.

We also measured the $(4550-4850)$\AA$/(7000-7300)$\AA\ flux ratio. Here the flux in the region 7000--7300\,\AA\ replaces the 4200--4500\,\AA\ region as a proxy for the Fe~{\sc ii} emission. This [Fe~{\sc ii}] line is near [Ni~{\sc ii}], and can also can become heavily contaminated by [Ca~{\sc ii}] (see e.g.\ \citealp{2015ApJ...814L...2F}). The feature is also weaker in comparison to the blue [Fe~{\sc ii}] line. We see there is more scatter in the measurements, even for multiple observations taken at similar times of the same SN, which is presumably due to calibration differences (the $(4550-4850)/(4200-4500)$ line ratio uses neighbouring lines, so for those measurements, differences in the calibration would have to be more extreme to produce a large scatter). However, the overall picture does not look too dissimilar to the other line ratio, where the SNe with the highest line ratio tend to be slower fading. A plateau of an approximately constant line ratio out to around 300\,d is not apparent for the $(4550-4850)$\AA$/(7000-7300)$\AA\ ratio. With the larger scatter, this would be more difficult to discern anyway, but it is possible the apparent plateau that seems to be present in the $(4550-4850)$\AA$/(4200-4500)$\AA\ ratio could be being influenced by weak permitted Fe~{\sc ii} lines still being present in the early part of the timeframe we show in Fig.~\ref{fig:ratio} (see e.g.\ \citealp{2016MNRAS.462..649B}), as this would have more of an effect on the blue lines.

We also examined the emission line around 5500\,\AA, which is due to be a blend of [Fe~{\sc ii}] and [Fe~{\sc iii}], and we measured the ratio to the [Fe~{\sc iii}] 4700\,\AA\ line. In comparison to the two previously discussed ratios, this ratio appeared more similar between SNe~Ia of differing $\Delta{m}_{15}(B)$ values, and any differences in the first 300--400 days appear to be largely lost in the scatter that we see for single SN with multiple spectra. This might be due to the ratio effectively being [Fe~{\sc iii}]/([Fe~{\sc ii}]+[Fe~{\sc iii}]), so the differences/evolution will be more subtle than that is seen for a more direct [Fe~{\sc iii}]/[Fe~{\sc ii}].

The fact there appears to be a general trend of lower [Fe~{\sc iii}]/[Fe~{\sc ii}] flux ratio being found in objects with higher $\Delta{m}_{15}(B)$ values, but there is a notable outlier at the very high end of the included $\Delta{m}_{15}(B)$ distribution is interesting. This could indicate that there exists a majority population of SNe~Ia that fall into a distribution where the ionisation of the central regions of the ejecta is positively correlated with the $^{56}$Ni yield of the explosion. In this scenario, the outlier of this distribution, SN\,2003hv, could have something fundamentally different about the makeup of the central regions of the ejecta, possibly related to the explosion mechanism. As previously suggested for SN\,2003hv \citep{2011MNRAS.416..881M}, an obvious difference could be a substantially lower density for the central portions of the ejecta in objects such as this, as lower density inhibits recombination and thus produces a higher ionisation balance. Another potential reason could be the degree to which $^{56}$Ni is mixed in the inner layers. \citet{2022A&A...660A..96B} have suggested that highly mixed inner regions could yield a higher ionisation balance of IGEs in the nebular phase. It would be worthwhile obtaining nebular spectroscopy of normal SNe~Ia with $\Delta{m}_{15}(B)$ values at the high end of the distribution, to see if the high ionisation seen in SN\,2003hv is common among SNe~Ia that have these high $\Delta{m}_{15}(B)$ values, or if SN\,2003hv is still an outlier even among these.

\section{Comparison to explosion models}
When considering the thermonuclear explosion of a carbon-oxygen white dwarf (WD), there are two ways the burning can propagate: in a super-sonic detonation or a sub-sonic deflagration. The nature of the burning products depend on the density of the material, with high densities leading to the WD material being completely burnt to IGEs. In a detonation, the material cannot expand prior to being reached by the burning front. Given the high densities of a $M_{\mathrm{Ch}}$ WD, the detonation of such a WD would lead to the ejecta being made up almost entirely of IGEs \citep{1971ApJ...165...87A}. As normal SNe~Ia synthesise a large amount of IMEs (e.g.\ \citealp{2007Sci...315..825M}, \citealp{2011MNRAS.410.1725T}), the detonation of a near-$M_{\mathrm{Ch}}$ WD is not the favoured mechanism for producing them. Instead there must be sufficient material at lower densities than those found in a near-$M_{\mathrm{Ch}}$ WD. The two currently favoured models are the detonation of a sub-$M_{\mathrm{Ch}}$ WD \citep{1994ApJ...423..371W,2010ApJ...714L..52S} or the delayed detonation of a near-$M_{\mathrm{Ch}}$ WD \citep{1991A&A...245..114K}. In the first scenario, the lower densities arise from the lower mass of the WD, and in the second scenario they come from the flame initially propagating as a sub-sonic deflagration (allowing some expansion of the WD), before a detonation is triggered. There are then several different scenarios within these two primary mechanisms, considering, for example, how the sub-$M_{\mathrm{Ch}}$ WD explosion is triggered or how the near-$M_{\mathrm{Ch}}$ deflagration-to-detonation transition is triggered.

Given that the luminosity of a SN~Ia is powered by the nuclear decay of $^{56}$Ni and its daughter isotope $^{56}$Co, the mass of $^{56}$Ni synthesised in the explosion and its distribution within the ejecta can produce much of the diversity seen in the normal SN~Ia population. Both of the two primary explosion mechanism candidates discussed above are plausibly able to reproduce many of the observables seen in SNe~Ia, including the range of peak brightnesses (e.g.\ \citealp{1995ApJ...444..831H}, \citealp{2010ApJ...714L..52S}, \citealp{2013MNRAS.429.2127B}, \citealp{2018A&A...614A.115M}, \citealp{2018ApJ...854...52S}). Shortly after explosion, different explosion models and progenitor configurations are often predicted to produce substantial differences in observable features, even if the model predictions have largely converged by the time the SN approaches peak brightness (e.g.\ \citealp{2017MNRAS.472.2787N}). Given how early our first spectrum and multi-colour photometry were taken after explosion, it is therefore worth comparing our data to some of these models.

\citet[][see also \citealp{2013MNRAS.429.2127B}]{2014MNRAS.441..532D} present a series of models exploring two types of delayed detonation. One was a `standard' delayed detonation (DDC), where the burning front initially propagates as a deflagration before transitioning to a detonation. Another was a pulsating delayed detonation (PDD). In this model, the burning again begins as a deflagration, but if the deflagration velocity is low, the flame can potentially be quenched, and the WD undergo a strong pulsation, triggering the detonation (see e.g.\ \citealp{1991A&A...245L..25K}). We compare one of each of these two $M_{\mathrm{Ch}}$ explosion mechanisms from \citet{2014MNRAS.441..532D} to our observations of SN\,2020nlb. The prompt detonation of a $\sim$1\,$M_{\odot}$ WD could potentially produce a similar amount of $^{56}$Ni that we estimate was synthesised in SN\,2020nlb \citep{2010ApJ...714L..52S,2014ApJ...785..105M,2018ApJ...854...52S}. As well as the $M_{\mathrm{Ch}}$ models from \citet{2014MNRAS.441..532D}, we also compare our observations of SN\,2020nlb to the prompt detonation of a $M_{\mathrm{WD}}=1.06$\,$M_{\odot}$ WD \citep{2010ApJ...714L..52S,2017MNRAS.472.2787N}.

\begin{figure}[ht]
    \centering
    \includegraphics[width=\columnwidth]{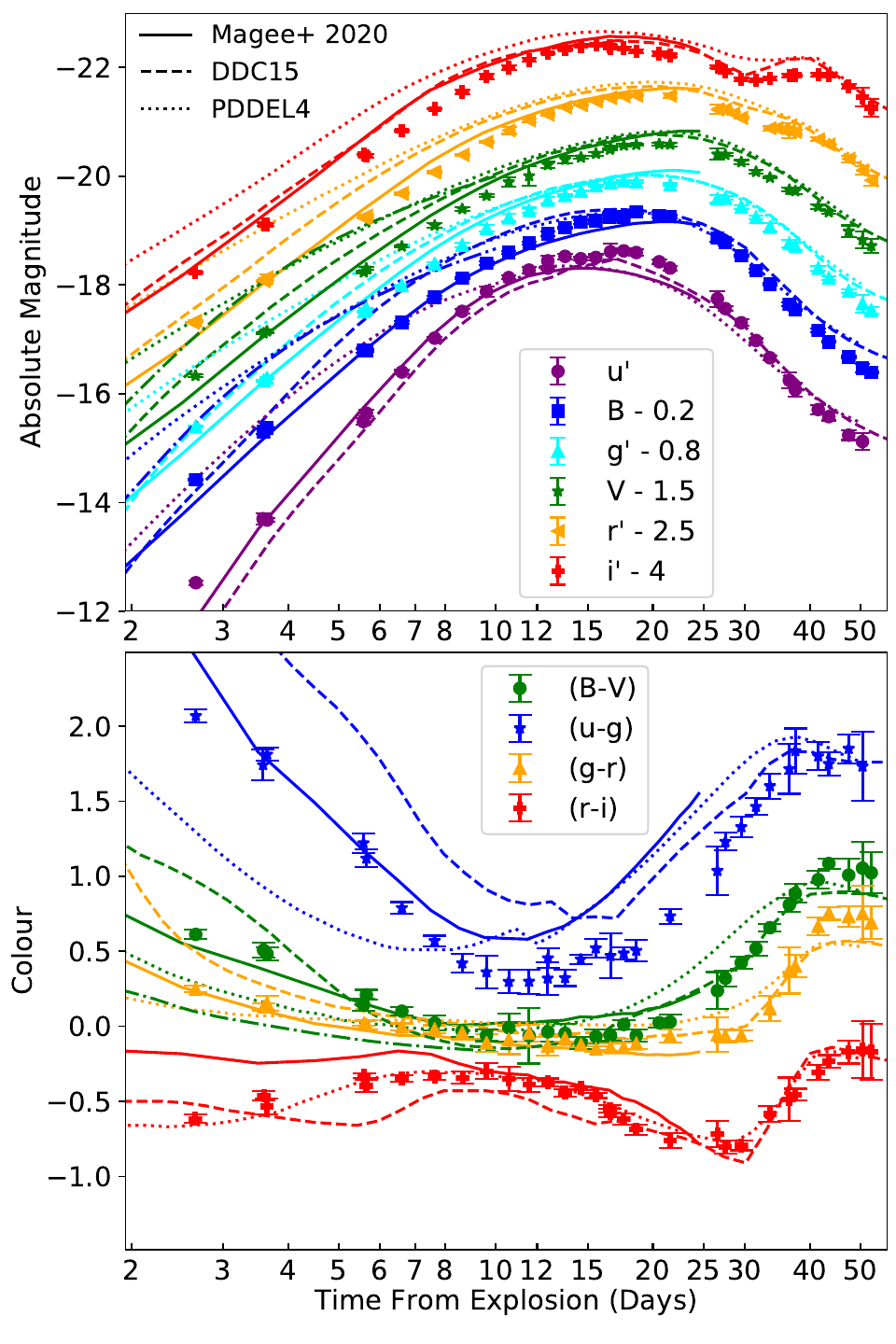}
    \caption{Light curve and colours of SN\,2020nlb (corrected for extinction) compared to models. The dashed and dotted lines represent the delayed detonation DDC15 and pulsating delayed detonation PDDEL4 models from \citet{2014MNRAS.441..532D} respectively. The solid lines show the EXP\_Ni0.6\_KE1.10\_P9.7 model from \citet{2020A&A...634A..37M}. The dot-dash lines show early $BV$ photometry for a $M_{\mathrm{WD}}=1.06$\,$M_{\odot}$ detonation \citep{2017MNRAS.472.2787N}. The time axis is shown in log space to make early-time variations clearer.}
    \label{fig:modslc}
\end{figure} 

We first compare to light curve models in Fig.~\ref{fig:modslc}. Here we used the DDC15 and PDDEL4 $M_{\mathrm{Ch}}$ model spectral sequences published by \citet{2013MNRAS.429.2127B} and \citet{2014MNRAS.441..532D}. The DDC15 and PDDEL4 models produced 0.511 and 0.529\,$M_{\odot}$ of $^{56}$Ni respectively \citep{2014MNRAS.441..532D}. These spectral sequences were then integrated using filter response curves that match our SN\,2020nlb observations. For the sub-$M_{\mathrm{Ch}}$ comparison, we use the early light curve of the $M_{\mathrm{WD}}=1.06$\,$M_{\odot}$ WD detonation from \citet{2017MNRAS.472.2787N}, which produced 0.56\,$M_{\odot}$ of $^{56}$Ni. In addition to these explosion models we also compared the early light curve and colours to the $^{56}$Ni distribution models from \citet{2020A&A...634A..37M}. We show the EXP\_Ni0.6\_KE1.10\_P9.7 model from \citet{2020A&A...634A..37M} in Fig.~\ref{fig:modslc}. This EXP\_Ni0.6\_KE1.10\_P9.7 model matches the colour evolution of SN\,2020nlb quite well. Note also that the $(u-g)$ comparison is only approximate. \citet{2020A&A...634A..37M} published $U$-band photometry for their models, rather than $u$-band, and the only correction we make here before plotting the model is to convert from the Vega to the AB magnitude system.

\begin{figure*}
    \centering
    \includegraphics[width=2\columnwidth]{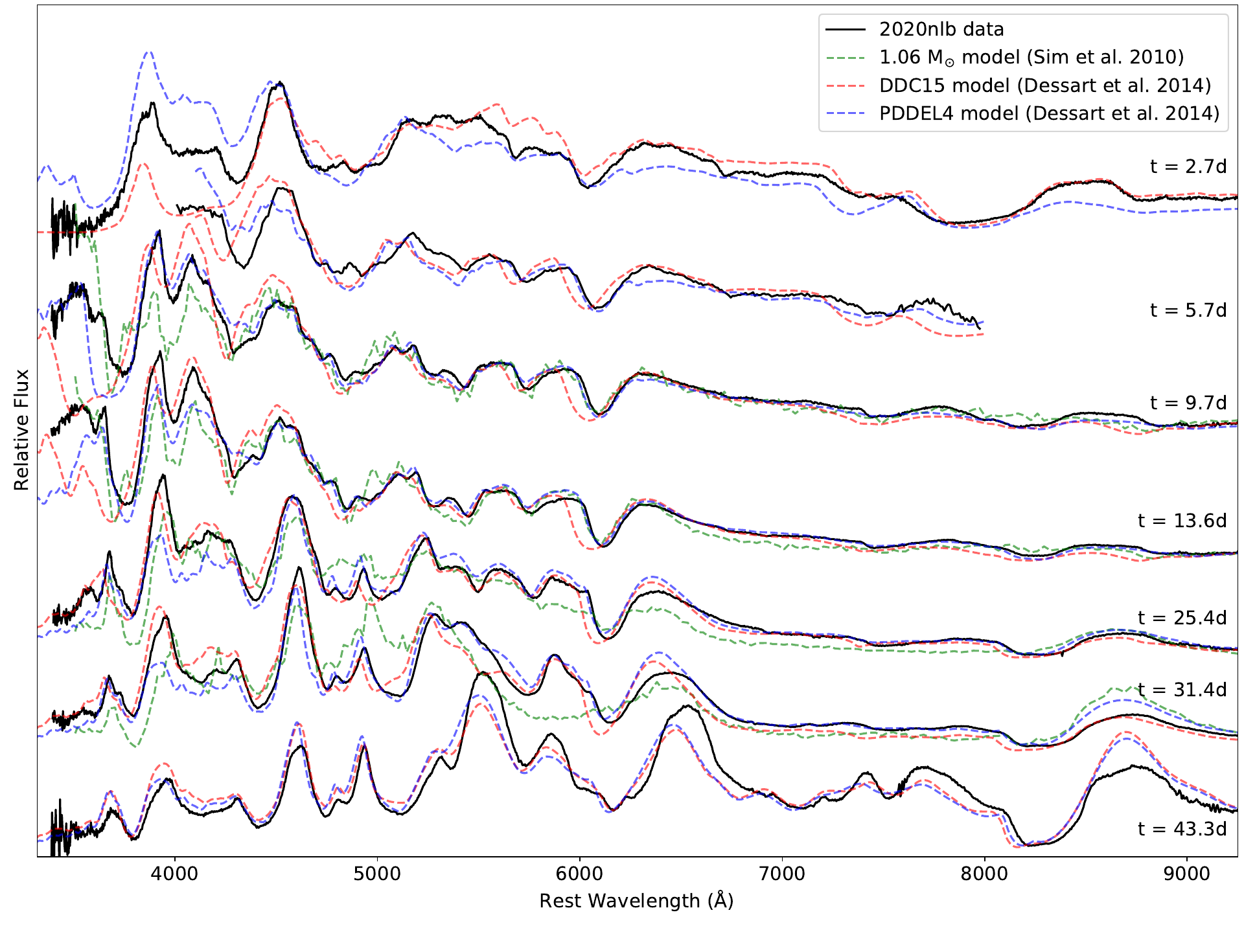}
    \caption{Comparison between selected spectra of SN\,2020nlb (in black) to various models. The red dashed lines show the `DDC15' delayed detonation model and the blue dashed lines show the `PDDEL4' pulsating delayed detonation model, both from \citet{2014MNRAS.441..532D}. The green lines show the $M_{\mathrm{WD}}=1.06$\,$M_{\odot}$ detonation model from \citet{2010ApJ...714L..52S}. The pre-maximum model spectral sequence of this model has high-cadence temporal sampling, so we co-add nearby epochs to increase the S/N. In these comparisons, we do not choose the best-matching spectra from the model sequence, but the spectrum taken nearest to the post-first light time of our data for SN\,2020nlb. The time after first light (rather than time with respect to peak) of each SN\,2020nlb spectrum is indicated.}
    \label{fig:modsspec}
\end{figure*}

In Fig.~\ref{fig:modsspec}, we compare select spectra of SN\,2020nlb to the DDC15 and PDDEL4 models published by \citet{2013MNRAS.429.2127B} and \citet{2014MNRAS.441..532D}. Given these models have only been matched to the approximate $^{56}$Ni yield of SN\,2020nlb, both the DDC and PDD models produce a fairly good match from the $t=-4$\,d spectrum onwards. The two models also produce reasonably similar spectra during these times. One noticeable difference is the Si~{\sc ii} 6355\,\AA\ line extending to higher velocities in the DDC model, whereas the line profile of the PDD model matches the SN\,2020nlb data well. \citet{2014MNRAS.441..532D} noted in their comparisons between their models and data, that while the DDC models produce a Si~{\sc ii} 6355\,\AA\ feature that well matches SNe~Ia with broader lines (see also \citealp{2015MNRAS.448.2766B}), it does not match well to SN\,2011fe. From the early light-curve fitting, we estimated that our first NOT spectrum of SN\,2020nlb was taken 2.7\,days after first light. At this epoch there are large differences between SN\,2020nlb and the two models, as well as the two models being very different from each other. It is worth noting that the time of first light we determined for SN\,2020nlb could be offset from the true explosion date: 1) if there was a significant dark phase, or 2) if there was a linear rise in the light curve at the earliest epochs. These two unknowns could potentially move the true explosion date earlier or later than our first light date respectively.

From the first comparison epoch it appears the PDD model is too blue and the DDC model has too much line blanketing at $<$\,4300\,\AA. Another substantial difference is the presence of strong C~{\sc ii} absorption in the PDD model, which is displayed neither by the DDC model or SN\,2020nlb. The $M_{\mathrm{WD}}=1.06$\,$M_{\odot}$ sub-$M_{\mathrm{Ch}}$ detonation model from \citet{2010ApJ...714L..52S} appears to match SN\,2020nlb quite well from around 1\,week prior to peak, but the spectral sequence for that model does not stretch to the earliest phases we have for SN\,2020nlb, precluding a comparison with the early data. 

\section{Conclusions and summary}
In this work we have presented observations of the nearby SN~Ia SN\,2020nlb, residing in the Virgo Cluster member M85. We briefly summarise the work below:

\begin{enumerate}
    \item We estimate that SN\,2020nlb was discovered just 2\,days after first light. The early-time light curve, beginning 0.7\,d after discovery, shows no clear evidence of a bump, and the early photometry in each optical filter is well described by a power-law rise. We find that fitting the data for each filter with a power-law rise yields $t_0$ values that agree well, indicating that cross-filter systematic uncertainties in derived $t_0$ values are low, with the possible exception of the $u'$ band.
    \item With $\Delta{m}_{15}(B)=1.28\pm0.02$\,mag, SN\,2020nlb declines faster than the average `normal' SN~Ia, implying it has a lower luminosity than the typical normal SN~Ia. We find it to be photometrically and spectroscopically similar to SN\,2015F.
    \item Our first spectrum was taken $16.1$\,d prior to \textit{B}-band maximum, 2.6\,d after first light, and shows strong features from singly ionised metals such as Fe~{\sc ii} and Ti~{\sc ii}, along with the usual Si~{\sc ii} features. It is very different to the early spectra of SN\,2011fe. No clear C~{\sc ii} 6580\,\AA\ line is detected at any epoch. 
    \item A nebular spectrum taken 594\,d after maximum light shows that the previously strong [Fe~{\sc iii}] emission line had disappeared, with the ionisation balance of the ejecta falling. This transition to lower ionisation appears to be already underway around 1\,yr after maximum light. 
    \item Comparing the flux ratio of the regions of the nebular spectrum dominated by [Fe~{\sc iii}] and [Fe~{\sc ii}] indicates that, among spectroscopically-normal SNe~Ia, more luminous objects tend to show higher ionisation in the nebular phase. There is however a notable outlier in SN\,2003hv.
    \item Using the multi-colour light curves of SN\,2020nlb, we estimate the distance modulus of M85 to be $\mu_0=30.99\pm0.19$\,mag, corresponding to a distance of $15.8^{+1.4}_{-1.3}$\,Mpc.
\end{enumerate}

\begin{acknowledgements}
We thank the anonymous referee for comments on this paper. This work is based on observations obtained with the Liverpool Telescope (LT), Nordic Optical Telescope (NOT), Galileo National Telescope (TNG), Very Large Telescope (VLT) and Schmidt Telescope. The LT is operated on the island of La Palma by Liverpool John Moores University in the Spanish Observatorio del Roque de los Muchachos (ORM) of the Instituto de Astrofisica de Canarias (IAC) with financial support from the UK Science and Technology Facilities Council. The NOT is owned in collaboration by the University of Turku and Aarhus University, and operated jointly by Aarhus University, the University of Turku and the University of Oslo, representing Denmark, Finland and Norway, the University of Iceland and Stockholm University at the Observatorio del Roque de los Muchachos, La Palma, Spain, of the Instituto de Astrofisica de Canarias. The NOT data presented here were obtained with ALFOSC, which is provided by the Instituto de Astrofisica de Andalucia (IAA) under a joint agreement with the University of Copenhagen and NOTSA. The TNG is operated on the island of La Palma by the Fundacio\'n Galileo Galilei of the INAF (Istituto Nazionale di Astrofisica) at the Spanish ORM of the IAC. The VLT observations were collected at the European Organisation for Astronomical Research in the Southern Hemisphere under ESO programme 108.22P1 (PI:~Williams). The Schmidt Telescope (Asiago, Italy) belongs to INAF-Osservatorio Astronomico di Padova. This work made use of WISeREP \citep{2012PASP..124..668Y}.

R.~Kotak acknowledges support from the Research Council of Finland (340613).

S.~Mattila acknowledges support from the Academy of Finland project 350458. 

P.~Lundqvist acknowledges support from the Swedish Research Council. 

A.~Fiore acknowledges the support by the State of Hesse within the Research Cluster ELEMENTS (Project ID 500/10.006).  

M.~D.~Stritzinger is funded by the Independent Research Fund Denmark (IRFD, grant number  10.46540/2032-00022B ). 

S.~Moran acknowledges support from the Magnus Ehrnrooth Foundation and the Vilho, Yrj\"{o}, and Kalle V\"{a}is\"{a}l\"{a} Foundation. 

I.~Salmaso acknowledges support by the doctoral grant funded by Istituto Nazionale di Astrofisica via the University of Padova and the Italian Ministry of Education, University and Research (MIUR) and by fundings from MIUR, PRIN 2017 (grant 20179ZF5KS).  

\end{acknowledgements}

\bibliographystyle{aa}

\begin{appendix}

\section{Possible tension with the power-law rise}\label{a1}

\begin{figure}[ht]
\centering
\includegraphics[width=\columnwidth]{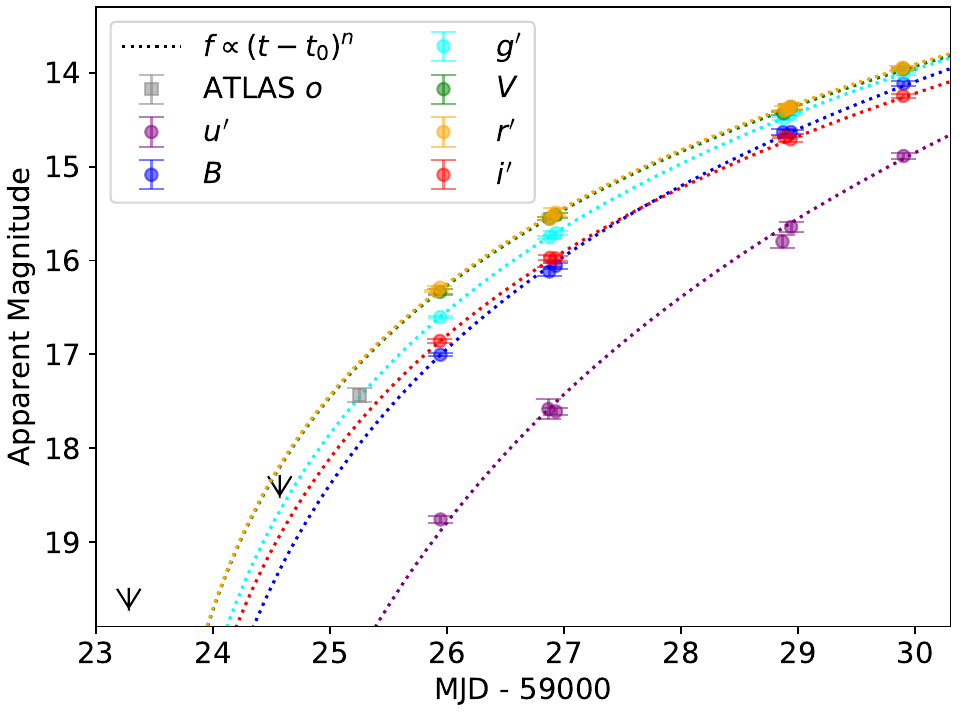}
\caption{Early light curve of SN\,2020nlb shown in apparent magnitude scale. The coloured dashed lines represent the power-law fits to the earlier data described in the main text. The earlier upper limit is in the ATLAS-$cyan$ filter, and the second later upper limit is from the Itagaki Astronomical Observatory’s 0.35 m telescope reported in \citet{2021ApJ...922...21S}. No offset has been applied to any of the photometric data here, so it is easier to visually interpret the upper limits, which are in different filters to the detections.}
\label{logearly}
\end{figure}

Fig.~\ref{logearly} shows the first few days of our multi-colour photometry and the power-law fits to the data, along with two non-detections. As can be seen from the figure, the earlier of the two non-detections, taken in the ATLAS-$cyan$ filter (approximately like a $g+r$ filter) does not really help constrain the earliest rise, as from the power-law fits, we would expect the SN to have been much fainter than that upper limit at that epoch. So that point only really rules out a strong early bump at that portion of the light curve.

The later of the two upper limits was taken using the Itagaki Astronomical Observatory’s 0.35 m telescope \citep{2021ApJ...922...21S}, and SN\,2020nlb was not detected down to a limiting magnitude of 18.5 at 59024.57\,MJD. Our power-law fits to our multi-colour early data, would predict the SN to be fainter than this in several filters, but brighter in the central region of the optical, with a \textit{V} and $r'$ magnitude of $\sim$\,18.2. This raises the possibility that the initial rise of the SN may have been more rapid than we estimate from the power-law fits, which if true, would in turn likely mean the real first-light date may have been slightly later than we estimate.

\section{Photometry}.
\centering
\tablefirsthead{\hline\hline MJD &$t$ [days] &Instrument &Filter &Magnitude\\\hline}
\tablehead{\hline\hline MJD &$t$ [days] &Instrument &Filter &Magnitude\\\hline}
\tablecaption{Optical photometry of SN\,2020nlb taken with the LT, NOT, and the 67/91 Schmidt Telescope.\vspace{-0.5cm}\label{tab:phot}}
\begin{supertabular}{lcccc}
\hline\hline
59025.94 & -16.12 & LT/IO:O & $u'$ & $18.76\pm0.04$\\ 
59026.87 & -15.20 & Schmidt & $u'$ & $17.58\pm0.10$\\ 
59026.93 & -15.13 & LT/IO:O & $u'$ & $17.61\pm0.04$\\ 
59028.87 & -13.20 & Schmidt & $u'$ & $15.80\pm0.07$\\ 
59028.94 & -13.13 & LT/IO:O & $u'$ & $15.64\pm0.06$\\ 
59029.90 & -12.17 & LT/IO:O & $u'$ & $14.89\pm0.03$\\ 
59030.93 & -11.14 & LT/IO:O & $u'$ & $14.26\pm0.03$\\ 
59031.92 & -10.16 & LT/IO:O & $u'$ & $13.77\pm0.04$\\ 
59032.90 & -9.18 & LT/IO:O & $u'$ & $13.41\pm0.09$\\ 
59033.89 & -8.19 & LT/IO:O & $u'$ & $13.15\pm0.03$\\ 
59034.89 & -7.20 & LT/IO:O & $u'$ & $13.01\pm0.06$\\ 
59035.89 & -6.20 & Schmidt & $u'$ & $12.84\pm0.09$\\ 
59035.90 & -6.19 & LT/IO:O & $u'$ & $12.98\pm0.03$\\ 
59036.89 & -5.20 & LT/IO:O & $u'$ & $12.76\pm0.05$\\ 
59037.86 & -4.23 & Schmidt & $u'$ & $12.81\pm0.03$\\ 
59038.86 & -3.23 & Schmidt & $u'$ & $12.78\pm0.06$\\ 
59039.91 & -2.19 & LT/IO:O & $u'$ & $12.68\pm0.15$\\ 
59040.91 & -1.19 & LT/IO:O & $u'$ & $12.66\pm0.04$\\ 
59041.91 & -0.19 & LT/IO:O & $u'$ & $12.69\pm0.06$\\ 
59043.90 & 1.79 & LT/IO:O & $u'$ & $12.86\pm0.04$\\ 
59044.92 & 2.81 & LT/IO:O & $u'$ & $12.98\pm0.04$\\ 
59049.94 & 7.82 & NOT/ALFOSC & $u'$ & $13.54\pm0.14$\\ 
59050.89 & 8.77 & LT/IO:O & $u'$ & $13.72\pm0.05$\\ 
59052.90 & 10.77 & LT/IO:O & $u'$ & $13.98\pm0.06$\\ 
59054.89 & 12.76 & LT/IO:O & $u'$ & $14.31\pm0.04$\\ 
59056.89 & 14.76 & LT/IO:O & $u'$ & $14.63\pm0.04$\\ 
59059.90 & 17.75 & LT/IO:O & $u'$ & $15.04\pm0.15$\\ 
59060.88 & 18.73 & LT/IO:O & $u'$ & $15.21\pm0.13$\\ 
59064.88 & 22.72 & LT/IO:O & $u'$ & $15.58\pm0.08$\\ 
59066.87 & 24.71 & LT/IO:O & $u'$ & $15.71\pm0.07$\\ 
59070.87 & 28.70 & LT/IO:O & $u'$ & $16.05\pm0.09$\\ 
59073.86 & 31.68 & LT/IO:O & $u'$ & $16.16\pm0.16$\\ 
59163.27 & 120.88 & LT/IO:O & $u'$ & $18.29\pm0.10$\\ 
59172.25 & 129.84 & LT/IO:O & $u'$ & $18.50\pm0.05$\\ 
59189.22 & 146.77 & LT/IO:O & $u'$ & $18.92\pm0.16$\\ 
59194.23 & 151.77 & LT/IO:O & $u'$ & $19.00\pm0.07$\\ 
59201.23 & 158.75 & LT/IO:O & $u'$ & $19.17\pm0.05$\\ 
59206.16 & 163.67 & LT/IO:O & $u'$ & $19.13\pm0.07$\\ 
59233.14 & 190.58 & NOT/ALFOSC & $u'$ & $19.65\pm0.07$\\ 
59257.12 & 214.51 & LT/IO:O & $u'$ & $20.13\pm0.09$\\ 
\hline
59025.94 & -16.12 & LT/IO:O & $g'$ & $16.60\pm0.01$\\ 
59026.88 & -15.19 & Schmidt & $g'$ & $15.75\pm0.02$\\ 
59026.93 & -15.14 & LT/IO:O & $g'$ & $15.71\pm0.02$\\ 
59028.88 & -13.19 & Schmidt & $g'$ & $14.49\pm0.01$\\ 
59028.94 & -13.13 & LT/IO:O & $g'$ & $14.43\pm0.02$\\ 
59029.90 & -12.17 & LT/IO:O & $g'$ & $14.01\pm0.02$\\ 
59030.93 & -11.14 & LT/IO:O & $g'$ & $13.61\pm0.02$\\ 
59031.92 & -10.16 & LT/IO:O & $g'$ & $13.26\pm0.04$\\ 
59032.90 & -9.18 & LT/IO:O & $g'$ & $12.96\pm0.06$\\ 
59033.89 & -8.19 & LT/IO:O & $g'$ & $12.76\pm0.07$\\ 
59034.89 & -7.20 & LT/IO:O & $g'$ & $12.63\pm0.05$\\ 
59035.89 & -6.20 & LT/IO:O & $g'$ & $12.43\pm0.06$\\ 
59036.89 & -5.20 & LT/IO:O & $g'$ & $12.35\pm0.02$\\ 
59037.88 & -4.21 & Schmidt & $g'$ & $12.27\pm0.01$\\
59038.88 & -3.22 & Schmidt & $g'$ & $12.17\pm0.02$\\ 
59039.91 & -2.19 & LT/IO:O & $g'$ & $12.12\pm0.02$\\ 
59040.91 & -1.19 & LT/IO:O & $g'$ & $12.08\pm0.02$\\ 
59041.91 & -0.19 & LT/IO:O & $g'$ & $12.10\pm0.03$\\ 
59044.92 & 2.81 & LT/IO:O & $g'$ & $12.15\pm0.02$\\ 
59049.94 & 7.82 & NOT/ALFOSC & $g'$ & $12.41\pm0.08$\\ 
59050.89 & 8.77 & LT/IO:O & $g'$ & $12.40\pm0.04$\\ 
59052.90 & 10.77 & LT/IO:O & $g'$ & $12.56\pm0.03$\\ 
59054.89 & 12.76 & LT/IO:O & $g'$ & $12.75\pm0.04$\\ 
59056.89 & 14.76 & LT/IO:O & $g'$ & $12.93\pm0.07$\\ 
59059.90 & 17.75 & LT/IO:O & $g'$ & $13.23\pm0.07$\\ 
59060.88 & 18.74 & LT/IO:O & $g'$ & $13.29\pm0.07$\\ 
59064.88 & 22.72 & LT/IO:O & $g'$ & $13.69\pm0.04$\\ 
59066.88 & 24.72 & LT/IO:O & $g'$ & $13.87\pm0.04$\\ 
59070.87 & 28.70 & LT/IO:O & $g'$ & $14.11\pm0.03$\\ 
59073.86 & 31.68 & LT/IO:O & $g'$ & $14.34\pm0.17$\\ 
59075.86 & 33.68 & LT/IO:O & $g'$ & $14.46\pm0.05$\\ 
59163.27 & 120.88 & LT/IO:O & $g'$ & $16.04\pm0.01$\\ 
59172.25 & 129.84 & LT/IO:O & $g'$ & $16.20\pm0.02$\\ 
59189.22 & 146.77 & LT/IO:O & $g'$ & $16.48\pm0.01$\\ 
59194.24 & 151.77 & LT/IO:O & $g'$ & $16.55\pm0.01$\\ 
59201.24 & 158.76 & LT/IO:O & $g'$ & $16.66\pm0.01$\\ 
59206.17 & 163.67 & LT/IO:O & $g'$ & $16.76\pm0.02$\\ 
59212.14 & 169.63 & LT/IO:O & $g'$ & $16.86\pm0.01$\\ 
59233.15 & 190.59 & NOT/ALFOSC & $g'$ & $17.09\pm0.03$\\ 
59257.13 & 214.51 & LT/IO:O & $g'$ & $17.52\pm0.01$\\ 
59262.08 & 219.46 & LT/IO:O & $g'$ & $17.65\pm0.05$\\ 
59268.04 & 225.40 & LT/IO:O & $g'$ & $17.69\pm0.02$\\ 
59291.10 & 248.40 & NOT/ALFOSC & $g'$ & $17.97\pm0.05$\\
59314.03 & 271.28 & NOT/ALFOSC & $g'$ & $18.29\pm0.03$\\ 
59322.92 & 280.15 & NOT/ALFOSC & $g'$ & $18.39\pm0.03$\\ 
59353.00 & 310.15 & NOT/ALFOSC & $g'$ & $18.88\pm0.02$\\ 
59367.92 & 325.04 & NOT/ALFOSC & $g'$ & $19.16\pm0.04$\\ 
59411.89 & 368.90 & LT/IO:O & $g'$ & $19.75\pm0.08$\\ 
\hline
59025.91 & -16.15 & LT/IO:O & $r'$ & $16.32\pm0.01$\\ 
59025.94 & -16.12 & LT/IO:O & $r'$ & $16.29\pm0.02$\\ 
59026.88 & -15.19 & Schmidt & $r'$ & $15.56\pm0.02$\\ 
59026.91 & -15.16 & LT/IO:O & $r'$ & $15.52\pm0.01$\\ 
59026.93 & -15.14 & LT/IO:O & $r'$ & $15.49\pm0.05$\\ 
59028.88 & -13.19 & Schmidt & $r'$ & $14.40\pm0.01$\\
59028.90 & -13.17 & LT/IO:O & $r'$ & $14.38\pm0.03$\\ 
59028.92 & -13.14 & LT/IO:O & $r'$ & $14.37\pm0.03$\\ 
59028.94 & -13.13 & LT/IO:O & $r'$ & $14.36\pm0.02$\\ 
59029.89 & -12.19 & LT/IO:O & $r'$ & $13.95\pm0.02$\\ 
59029.90 & -12.17 & LT/IO:O & $r'$ & $13.95\pm0.01$\\ 
59030.90 & -11.17 & LT/IO:O & $r'$ & $13.56\pm0.02$\\ 
59030.93 & -11.14 & LT/IO:O & $r'$ & $13.56\pm0.02$\\ 
59031.94 & -10.14 & LT/IO:O & $r'$ & $13.24\pm0.01$\\ 
59032.90 & -9.18 & LT/IO:O & $r'$ & $13.00\pm0.02$\\ 
59033.89 & -8.19 & LT/IO:O & $r'$ & $12.79\pm0.07$\\ 
59034.89 & -7.20 & LT/IO:O & $r'$ & $12.61\pm0.03$\\ 
59035.89 & -6.20 & LT/IO:O & $r'$ & $12.50\pm0.02$\\ 
59036.89 & -5.20 & LT/IO:O & $r'$ & $12.36\pm0.01$\\
59037.88 & -4.21 & Schmidt & $r'$ & $12.33\pm0.01$\\ 
59038.88 & -3.21 & Schmidt & $r'$ & $12.26\pm0.01$\\ 
59039.86 & -2.23 & Schmidt & $r'$ & $12.19\pm0.02$\\ 
59039.90 & -2.20 & LT/IO:O & $r'$ & $12.19\pm0.02$\\ 
59040.91 & -1.18 & LT/IO:O & $r'$ & $12.16\pm0.03$\\ 
59041.89 & -0.21 & LT/IO:O & $r'$ & $12.15\pm0.02$\\ 
59044.92 & 2.81 & LT/IO:O & $r'$ & $12.15\pm0.04$\\ 
59049.94 & 7.82 & NOT/ALFOSC & $r'$ & $12.40\pm0.08$\\ 
59050.89 & 8.77 & LT/IO:O & $r'$ & $12.40\pm0.02$\\ 
59051.90 & 9.77 & LT/IO:O & $r'$ & $12.48\pm0.02$\\ 
59052.90 & 10.77 & LT/IO:O & $r'$ & $12.56\pm0.01$\\ 
59056.90 & 14.76 & LT/IO:O & $r'$ & $12.75\pm0.04$\\ 
59057.89 & 15.75 & LT/IO:O & $r'$ & $12.75\pm0.03$\\ 
59058.88 & 16.74 & LT/IO:O & $r'$ & $12.78\pm0.02$\\ 
59059.86 & 17.72 & LT/IO:O & $r'$ & $12.79\pm0.14$\\ 
59060.88 & 18.74 & LT/IO:O & $r'$ & $12.82\pm0.04$\\ 
59064.88 & 22.72 & LT/IO:O & $r'$ & $12.96\pm0.03$\\ 
59066.88 & 24.72 & LT/IO:O & $r'$ & $13.05\pm0.02$\\ 
59070.87 & 28.70 & LT/IO:O & $r'$ & $13.31\pm0.06$\\ 
59073.86 & 31.68 & LT/IO:O & $r'$ & $13.53\pm0.08$\\ 
59075.86 & 33.68 & LT/IO:O & $r'$ & $13.71\pm0.10$\\ 
59163.28 & 120.88 & LT/IO:O & $r'$ & $16.54\pm0.01$\\ 
59172.25 & 129.84 & LT/IO:O & $r'$ & $16.80\pm0.02$\\ 
59189.22 & 146.77 & LT/IO:O & $r'$ & $17.25\pm0.02$\\ 
59194.24 & 151.77 & LT/IO:O & $r'$ & $17.37\pm0.03$\\ 
59201.24 & 158.76 & LT/IO:O & $r'$ & $17.54\pm0.01$\\ 
59206.17 & 163.68 & LT/IO:O & $r'$ & $17.66\pm0.03$\\ 
59212.14 & 169.63 & LT/IO:O & $r'$ & $17.81\pm0.02$\\
59225.15 & 182.61 & Schmidt & $r'$ & $18.24\pm0.08$\\
59233.15 & 190.59 & NOT/ALFOSC & $r'$ & $18.21\pm0.03$\\ 
59257.13 & 214.51 & LT/IO:O & $r'$ & $18.81\pm0.02$\\ 
59262.08 & 219.46 & LT/IO:O & $r'$ & $19.08\pm0.15$\\ 
59268.04 & 225.40 & LT/IO:O & $r'$ & $19.05\pm0.03$\\ 
59291.10 & 248.40 & NOT/ALFOSC & $r'$ & $19.43\pm0.03$\\ 
59314.03 & 271.28 & NOT/ALFOSC & $r'$ & $19.77\pm0.03$\\ 
59322.93 & 280.15 & NOT/ALFOSC & $r'$ & $19.89\pm0.04$\\ 
59353.00 & 310.15 & NOT/ALFOSC & $r'$ & $20.39\pm0.04$\\ 
59367.92 & 325.04 & NOT/ALFOSC & $r'$ & $20.77\pm0.09$\\ 
59411.89 & 368.90 & LT/IO:O & $r'$ & $21.33\pm0.34$\\ 
\hline
59025.94 & -16.12 & LT/IO:O & $i'$ & $16.86\pm0.02$\\ 
59026.88 & -15.18 & Schmidt& $i'$ & $15.97\pm0.03$\\
59026.93 & -15.14 & LT/IO:O & $i'$ & $15.98\pm0.03$\\ 
59028.89 & -13.18 & Schmidt & $i'$ & $14.68\pm0.01$\\ 
59028.93 & -13.13 & LT/IO:O & $i'$ & $14.71\pm0.03$\\ 
59029.90 & -12.17 & LT/IO:O & $i'$ & $14.25\pm0.02$\\ 
59030.93 & -11.14 & LT/IO:O & $i'$ & $13.85\pm0.01$\\ 
59031.92 & -10.15 & LT/IO:O & $i'$ & $13.53\pm0.04$\\ 
59032.90 & -9.18 & LT/IO:O & $i'$ & $13.25\pm0.05$\\ 
59033.89 & -8.19 & LT/IO:O & $i'$ & $13.09\pm0.05$\\ 
59034.89 & -7.19 & LT/IO:O & $i'$ & $12.95\pm0.03$\\ 
59035.89 & -6.20 & LT/IO:O & $i'$ & $12.83\pm0.02$\\ 
59036.89 & -5.20 & LT/IO:O & $i'$ & $12.76\pm0.02$\\ 
59037.88 & -4.21 & Schmidt & $i'$ & $12.69\pm0.01$\\ 
59038.88 & -3.21 & Schmidt & $i'$ & $12.67\pm0.01$\\
59039.87 & -2.23 & Schmidt & $i'$ & $12.69\pm0.02$\\ 
59039.91 & -2.19 & LT/IO:O & $i'$ & $12.73\pm0.02$\\ 
59040.91 & -1.18 & LT/IO:O & $i'$ & $12.73\pm0.02$\\ 
59041.91 & -0.19 & LT/IO:O & $i'$ & $12.79\pm0.03$\\ 
59043.90 & 1.79 & LT/IO:O & $i'$ & $12.83\pm0.04$\\ 
59044.92 & 2.81 & LT/IO:O & $i'$ & $12.87\pm0.03$\\ 
59049.94 & 7.82 & NOT/ALFOSC & $i'$ & $13.08\pm0.02$\\ 
59050.89 & 8.77 & LT/IO:O & $i'$ & $13.16\pm0.03$\\ 
59052.90 & 10.77 & LT/IO:O & $i'$ & $13.31\pm0.03$\\ 
59054.89 & 12.76 & LT/IO:O & $i'$ & $13.32\pm0.04$\\ 
59056.90 & 14.76 & LT/IO:O & $i'$ & $13.29\pm0.04$\\ 
59059.90 & 17.76 & LT/IO:O & $i'$ & $13.23\pm0.03$\\ 
59060.88 & 18.74 & LT/IO:O & $i'$ & $13.23\pm0.02$\\ 
59064.88 & 22.73 & LT/IO:O & $i'$ & $13.22\pm0.03$\\ 
59066.88 & 24.72 & LT/IO:O & $i'$ & $13.24\pm0.04$\\ 
59070.87 & 28.70 & LT/IO:O & $i'$ & $13.44\pm0.05$\\ 
59073.86 & 31.68 & LT/IO:O & $i'$ & $13.63\pm0.17$\\ 
59075.86 & 33.68 & LT/IO:O & $i'$ & $13.83\pm0.16$\\ 
59163.28 & 120.89 & LT/IO:O & $i'$ & $16.77\pm0.01$\\ 
59172.25 & 129.84 & LT/IO:O & $i'$ & $17.00\pm0.01$\\ 
59189.22 & 146.77 & LT/IO:O & $i'$ & $17.32\pm0.02$\\ 
59194.24 & 151.78 & LT/IO:O & $i'$ & $17.47\pm0.03$\\ 
59201.24 & 158.76 & LT/IO:O & $i'$ & $17.55\pm0.02$\\ 
59206.17 & 163.68 & LT/IO:O & $i'$ & $17.61\pm0.02$\\ 
59212.14 & 169.63 & LT/IO:O & $i'$ & $17.75\pm0.02$\\
59233.15 & 190.59 & NOT/ALFOSC & $i'$ & $18.08\pm0.03$\\  
59257.13 & 214.51 & LT/IO:O & $i'$ & $18.38\pm0.02$\\ 
59262.08 & 219.46 & LT/IO:O & $i'$ & $18.34\pm0.12$\\ 
59268.04 & 225.40 & LT/IO:O & $i'$ & $18.49\pm0.02$\\ 
59291.10 & 248.41 & NOT/ALFOSC & $i'$ & $18.87\pm0.03$\\ 
59314.03 & 271.28 & NOT/ALFOSC & $i'$ & $19.11\pm0.05$\\ 
59322.93 & 280.15 & NOT/ALFOSC & $i'$ & $19.23\pm0.06$\\ 
59353.00 & 310.16 & NOT/ALFOSC & $i'$ & $19.61\pm0.04$\\ 
59367.92 & 325.04 & NOT/ALFOSC & $i'$ & $19.87\pm0.06$\\ 
59411.89 & 368.90 & LT/IO:O & $i'$ & $20.42\pm0.19$\\ 
\hline
59025.94 & -16.12 & LT/IO:O & $B$ & $17.00\pm0.02$\\ 
59026.87 & -15.19 & Schmidt & $B$ & $16.12\pm0.05$\\ 
59026.93 & -15.13 & LT/IO:O & $B$ & $16.06\pm0.03$\\ 
59028.87 & -13.20 & Schmidt & $B$ & $14.63\pm0.03$\\ 
59028.94 & -13.13 & LT/IO:O & $B$ & $14.63\pm0.02$\\ 
59029.90 & -12.17 & LT/IO:O & $B$ & $14.11\pm0.03$\\ 
59030.93 & -11.14 & LT/IO:O & $B$ & $13.65\pm0.03$\\ 
59031.92 & -10.16 & LT/IO:O & $B$ & $13.30\pm0.02$\\ 
59032.90 & -9.18 & LT/IO:O & $B$ & $13.02\pm0.03$\\ 
59033.89 & -8.19 & LT/IO:O & $B$ & $12.82\pm0.03$\\ 
59034.89 & -7.20 & LT/IO:O & $B$ & $12.66\pm0.02$\\ 
59035.89 & -6.20 & LT/IO:O & $B$ & $12.48\pm0.04$\\ 
59036.89 & -5.20 & LT/IO:O & $B$ & $12.37\pm0.02$\\
59037.87 & -4.22 & Schmidt & $B$ & $12.27\pm0.03$\\
59038.87 & -3.22 & Schmidt & $B$ & $12.24\pm0.03$\\
59039.86 & -2.23 & Schmidt & $B$ & $12.18\pm0.03$\\ 
59039.91 & -2.19 & LT/IO:O & $B$ & $12.14\pm0.02$\\ 
59040.91 & -1.19 & LT/IO:O & $B$ & $12.17\pm0.02$\\ 
59041.91 & -0.19 & LT/IO:O & $B$ & $12.08\pm0.03$\\ 
59043.90 & 1.79 & LT/IO:O & $B$ & $12.16\pm0.02$\\ 
59044.92 & 2.81 & LT/IO:O & $B$ & $12.17\pm0.03$\\ 
59049.94 & 7.82 & NOT/ALFOSC & $B$ & $12.56\pm0.07$\\ 
59050.89 & 8.77 & LT/IO:O & $B$ & $12.65\pm0.04$\\ 
59052.90 & 10.77 & LT/IO:O & $B$ & $12.89\pm0.02$\\ 
59054.89 & 12.76 & LT/IO:O & $B$ & $13.16\pm0.03$\\ 
59056.89 & 14.76 & LT/IO:O & $B$ & $13.41\pm0.02$\\ 
59059.90 & 17.75 & LT/IO:O & $B$ & $13.78\pm0.03$\\ 
59060.88 & 18.73 & LT/IO:O & $B$ & $13.88\pm0.02$\\ 
59064.88 & 22.72 & LT/IO:O & $B$ & $14.26\pm0.04$\\ 
59066.87 & 24.72 & LT/IO:O & $B$ & $14.47\pm0.02$\\ 
59070.87 & 28.70 & LT/IO:O & $B$ & $14.74\pm0.02$\\ 
59073.86 & 31.68 & LT/IO:O & $B$ & $14.95\pm0.07$\\ 
59075.86 & 33.68 & LT/IO:O & $B$ & $15.03\pm0.02$\\ 
59163.27 & 120.88 & LT/IO:O & $B$ & $16.44\pm0.02$\\ 
59172.25 & 129.84 & LT/IO:O & $B$ & $16.59\pm0.02$\\ 
59189.22 & 146.77 & LT/IO:O & $B$ & $16.82\pm0.02$\\ 
59194.24 & 151.77 & LT/IO:O & $B$ & $16.95\pm0.02$\\ 
59201.24 & 158.76 & LT/IO:O & $B$ & $17.08\pm0.03$\\ 
59206.16 & 163.67 & LT/IO:O & $B$ & $17.12\pm0.03$\\ 
59212.14 & 169.63 & LT/IO:O & $B$ & $17.26\pm0.03$\\
59225.14 & 182.60 & Schmidt & $B$ & $17.50\pm0.03$\\
59233.14 & 190.59 & NOT/ALFOSC & $B$ & $17.58\pm0.02$\\ 
59257.13 & 214.51 & LT/IO:O & $B$ & $17.94\pm0.03$\\ 
59262.08 & 219.45 & LT/IO:O & $B$ & $17.94\pm0.06$\\ 
59268.04 & 225.40 & LT/IO:O & $B$ & $18.10\pm0.04$\\ 
59291.10 & 248.40 & NOT/ALFOSC & $B$ & $18.49\pm0.02$\\ 
59314.03 & 271.28 & NOT/ALFOSC & $B$ & $18.80\pm0.01$\\
59322.92 & 280.15 & NOT/ALFOSC & $B$ & $18.89\pm0.02$\\ 
59353.00 & 310.15 & NOT/ALFOSC & $B$ & $19.34\pm0.03$\\ 
59367.92 & 325.04 & NOT/ALFOSC & $B$ & $19.56\pm0.03$\\ 
59411.88 & 368.90 & LT/IO:O & $B$ & $20.45\pm0.18$\\  
\hline
59025.94 & -16.12 & LT/IO:O & $V$ & $16.34\pm0.03$\\ 
59026.87 & -15.19 & Schmidt & $V$ & $15.55\pm0.02$\\ 
59026.93 & -15.14 & LT/IO:O & $V$ & $15.52\pm0.03$\\ 
59028.87 & -13.19 & Schmidt & $V$ & $14.43\pm0.02$\\ 
59028.94 & -13.13 & LT/IO:O & $V$ & $14.36\pm0.03$\\ 
59029.90 & -12.17 & LT/IO:O & $V$ & $13.96\pm0.01$\\ 
59030.93 & -11.14 & LT/IO:O & $V$ & $13.57\pm0.04$\\ 
59031.92 & -10.16 & LT/IO:O & $V$ & $13.27\pm0.04$\\ 
59032.90 & -9.18 & LT/IO:O & $V$ & $13.03\pm0.02$\\ 
59033.89 & -8.19 & LT/IO:O & $V$ & $12.78\pm0.09$\\ 
59034.89 & -7.20 & LT/IO:O & $V$ & $12.66\pm0.18$\\ 
59035.89 & -6.20 & LT/IO:O & $V$ & $12.46\pm0.06$\\ 
59036.89 & -5.20 & LT/IO:O & $V$ & $12.36\pm0.06$\\ 
59037.87 & -4.22 & Schmidt & $V$ & $12.32\pm0.03$\\ 
59038.87 & -3.22 & Schmidt & $V$ & $12.25\pm0.03$\\ 
59039.86 & -2.23 & Schmidt & $V$ & $12.18\pm0.02$\\ 
59039.91 & -2.19 & LT/IO:O & $V$ & $12.15\pm0.04$\\ 
59040.91 & -1.19 & LT/IO:O & $V$ & $12.10\pm0.02$\\ 
59041.91 & -0.19 & LT/IO:O & $V$ & $12.09\pm0.02$\\ 
59043.90 & 1.79 & LT/IO:O & $V$ & $12.08\pm0.02$\\ 
59044.92 & 2.81 & LT/IO:O & $V$ & $12.08\pm0.04$\\ 
59049.94 & 7.82 & NOT/ALFOSC & $V$ & $12.27\pm0.10$\\ 
59050.89 & 8.77 & LT/IO:O & $V$ & $12.27\pm0.03$\\ 
59052.90 & 10.77 & LT/IO:O & $V$ & $12.41\pm0.04$\\ 
59054.89 & 12.76 & LT/IO:O & $V$ & $12.59\pm0.03$\\ 
59056.90 & 14.76 & LT/IO:O & $V$ & $12.69\pm0.02$\\ 
59059.90 & 17.76 & LT/IO:O & $V$ & $12.91\pm0.03$\\ 
59060.88 & 18.74 & LT/IO:O & $V$ & $12.94\pm0.06$\\ 
59064.88 & 22.72 & LT/IO:O & $V$ & $13.23\pm0.04$\\ 
59066.88 & 24.72 & LT/IO:O & $V$ & $13.33\pm0.03$\\ 
59070.87 & 28.70 & LT/IO:O & $V$ & $13.68\pm0.11$\\ 
59073.86 & 31.68 & LT/IO:O & $V$ & $13.84\pm0.16$\\ 
59075.86 & 33.68 & LT/IO:O & $V$ & $13.95\pm0.13$\\ 
59163.27 & 120.88 & LT/IO:O & $V$ & $16.12\pm0.01$\\ 
59172.25 & 129.84 & LT/IO:O & $V$ & $16.31\pm0.01$\\ 
59189.22 & 146.77 & LT/IO:O & $V$ & $16.61\pm0.02$\\ 
59194.24 & 151.77 & LT/IO:O & $V$ & $16.70\pm0.02$\\ 
59201.24 & 158.76 & LT/IO:O & $V$ & $16.82\pm0.02$\\ 
59206.17 & 163.67 & LT/IO:O & $V$ & $16.92\pm0.02$\\ 
59212.14 & 169.63 & LT/IO:O & $V$ & $17.05\pm0.03$\\ 
59225.15 & 182.61 & Schmidt & $V$ & $17.33\pm0.02$\\ 
59233.14 & 190.59 & NOT/ALFOSC & $V$ & $17.41\pm0.03$\\ 
59257.13 & 214.51 & LT/IO:O & $V$ & $17.75\pm0.02$\\ 
59262.08 & 219.45 & LT/IO:O & $V$ & $17.87\pm0.08$\\ 
59268.04 & 225.40 & LT/IO:O & $V$ & $17.89\pm0.03$\\ 
59291.10 & 248.40 & NOT/ALFOSC & $V$ & $18.37\pm0.03$\\ 
59314.03 & 271.28 & NOT/ALFOSC & $V$ & $18.66\pm0.03$\\ 
59322.92 & 280.15 & NOT/ALFOSC & $V$ & $18.75\pm0.02$\\ 
59353.00 & 310.15 & NOT/ALFOSC & $V$ & $19.31\pm0.07$\\ 
59367.92 & 325.04 & NOT/ALFOSC & $V$ & $19.48\pm0.04$\\ 
59411.89 & 368.90 & LT/IO:O & $V$ & $20.06\pm0.16$\\ 
\hline\hline
\end{supertabular}

\vspace{-6cm}

\begin{justify}
{\small
    \textbf{Note.} The epoch in days with respect to peak \textit{B}-band brightness (\textit{t}) is quoted in the rest frame.}
\end{justify}

\pagebreak

\section{Log of spectra}

\begin{table}[ht]
\caption{Log of Liverpool Telescope (LT), Nordic Optical Telescope (NOT), Galileo National Telescope (TNG), and Very Large Telescope (VLT) spectra taken of SN\,2020nlb.} \label{tab:log}      
\centering
\begin{tabular}{l c c c}          
\hline\hline                        
Instrument &Date [UT] &$t$ [days] &Exposure [s]\\
\hline                                   
LT SPRAT &2020 Jun 25.92 &$-16.1$ &1800\\
NOT ALFOSC &2020 Jun 25.97 &$-16.1$ &3600\\
LT SPRAT &2020 Jun 26.92 &$-15.1$ &1200\\
LT SPRAT &2020 Jun 28.93 &$-13.1$ &600\\
LT SPRAT &2020 Jun 29.89 &$-12.2$ &400\\
NOT ALFOSC &2020 Jul 2.95 &$-9.1$ &400\\
NOT ALFOSC &2020 Jul 6.91 &$-5.2$ &400\\
NOT ALFOSC &2020 Jul 7.89 &$-4.2$ &400\\
LT SPRAT &2020 Jul 9.90 &$-2.2$ &200\\
LT SPRAT &2020 Jul 13.89 &$+1.8$ &200\\
NOT ALFOSC &2020 Jul 18.89 &$+6.8$ &200\\
LT SPRAT &2020 Jul 22.91 &$+10.8$ &300\\
NOT ALFOSC &2020 Jul 24.89 &$+12.8$ &600\\
NOT ALFOSC &2020 Jul 30.89 &$+18.8$ &400\\
NOT ALFOSC &2020 Dec 12.26 &$+153$ &900\\
TNG LRS &2020 Dec 18.21 &$+159$ &900\\
NOT ALFOSC &2021 Jan 2.26 &$+174$ &1200\\
NOT ALFOSC &2021 Jan 19.13 &$+191$ &1200\\
NOT ALFOSC &2021 Feb 15.28 &$+218$ &1200\\
TNS LRS &2021 Mar 17.97 &$+248$ &900\\
NOT ALFOSC &2021 Mar 18.08 &$+248$ &2100\\
NOT ALFOSC &2021 Apr 21.08 &$+282$ &2100\\
TNG LRS &2020 May 6.08 &$+297$ &3600\\
NOT ALFOSC &2021 Jun 2.94 &$+325$ &2100\\
TNG LRS &2021 Jul 5.94 &$+360$ &2700\\
VLT FORS2 &2022 Feb 27.26 &$+594$ &2460\\
\hline
\end{tabular}
\begin{justify}
{\small
\textbf{Note.} The epoch in days with respect to peak \textit{B}-band brightness (\textit{t}) is quoted in the rest frame.}
\end{justify}
\end{table}

\section{Nebular spectra of SNe~Ia}

\begin{table*}[!ht]
    \centering
    \caption{Parameters of SNe~Ia with nebular spectra used in Fig.~\ref{fig:ratio}}
    \begin{tabular}{lcccccc}
    \hline\hline
        Supernova &$E(B-V)_{\mathrm{MW}}$ &$E(B-V)_{\mathrm{Host}}$ &$R_V$ (host) &$\Delta{m}_{15}(B)$ &References &Spectra References\\
        \hline
         SN\,1990N &0.022 & 0.00 & ... &1.03 &(1) &(2, 3)\\
         SN\,1994ae &0.026 &0.15 &3.1 &0.90 & (4) &(2, 5)\\
         SN\,1998bu &0.022 &0.32 &3.1 &1.02 & (6) & (2, 7, 8)\\
         SN\,2002er &0.138 &0.22 &3.1 &1.33 &(9) & (10)\\
         SN\,2003du &0.008 &0.00 & ... &1.02 & (11) & (12)\\
         SN\,2003hv &0.013 &0.00 & ... &1.61 &(13) & (13)\\
         SN\,2004eo &0.093 &0.00 & ... &1.46 &(14) & (14)\\
         SN\,2005cf &0.084 &0.10 & 3.1 &1.05 &(15) & (15)\\
         SN\,2007af &0.034 &0.13 & 2.98 &1.22 &(16, 17) &(2, 5)\\
         SN\,2007gi &0.019 &0.20 & 1.56 &1.33 &(18) & (18)\\
         SN\,2009ig &0.027 &0.00 &... &0.89 &(19) & (20)\\
         SN\,2011by &0.012 &0.053 &3.1 &1.14 &(21) & (22)\\
         SN\,2011fe &0.008 &0.014 &3.1 &1.11 &(23, 24) & (20, 25, 26, 27)\\
         SN\,2012cg &0.018 &0.18 &3.1 &0.86 &(28) & (29)\\
         SN\,2012ht &0.025 &0.00 &... &1.30 &(30) & (29)\\
         SN\,2012fr &0.018 &0.00 & ... &0.82 &(31) & (29, 32, 33)\\
         SN\,2013aa &0.146 &0.00 & ... &0.95 & (34) & (29, 32, 33, 35)\\
         SN\,2013cs &0.079 &0.00 & ... &0.81 & (36) & (29, 33)\\
         SN\,2013dy &0.132 &0.206 &3.1 &0.89 &(37) & (38, 39)\\
         SN\,2013gy &0.049 &0.11 &3.1 &1.23 &(40) & (32, 33)\\
         SN\,2015F &0.175 &0.085 &3.1 &1.35 &(41) & (33)\\
         SN\,2020nlb &0.026 &0.05 &3.1 &1.28 &(42) & (42)\\
    \hline\hline\\
    \end{tabular}
    \begin{justify}
    {\small
\textbf{Notes}. All Milky Way reddening values are from \citet{2011ApJ...737..103S}, and we assume $R_V=3.1$. A
    host $R_V$ value of 3.1 has been assumed where no $R_V$ was derived to accompany the published host reddening value. {SN Specific References:} 
    (1)~\citealp{1998AJ....115..234L};
    (2)~\citealp{2012MNRAS.425.1789S};
    (3)~\citealp{1998AJ....115.1096G};
    (4)~\citealp{2006A&A...460..793S};
    (5)~\citealp{2012AJ....143..126B};
    (6)~\citealp{1999ApJS..125...73J};
    (7)~\citealp{2008AJ....135.1598M};
    (8)~\citealp{2001ApJ...549L.215C};
    (9)~\citealp{2004MNRAS.355..178P}, who estimated total reddening of $E(B-V)=0.36$;
    (10)~\citealp{2005A&A...436.1021K};
    (11)~\citealp{2005A&A...429..667A};
    (12)~\citealp{2007A&A...469..645S};
    (13)~\citealp{2009A&A...505..265L};
    (14)~\citealp{2007MNRAS.377.1531P};
    (15)~\citealp{2009ApJ...697..380W};
    (16)~\citealp{2007ApJ...671L..25S};
    (17)~\citealp{2010ApJS..190..418G};
    (18)~\citealp{2010PASP..122....1Z};
    (19)~\citealp{2012ApJ...744...38F};
    (20)~\citealp{2020MNRAS.492.4325S};
    (21)~\citealp{2020MNRAS.491.5991F}, they derived SN\,2011by to have $E(B-V)_{\mathrm{Host}}$ of 0.039\,mag more than SN\,2011fe; 
    (22)~\citealp{2013MNRAS.430.1030S};
    (23)~\citealp{2013A&A...549A..62P};
    (24)~\citealp{2013NewA...20...30M};
    (25)~\citealp{2016ApJ...820...67Z};
    (26)~\citealp{2015MNRAS.450.2631M};
    (27)~\citealp{2022ApJ...926L..25T};
    (28)~\citealp{2016ApJ...820...92M};
    (29)~\citealp{2018MNRAS.477.3567M};
    (30)~\citealp{2015A&A...578A...9H};
    (31)~\citealp{2018ApJ...859...24C};
    (32)~\citealp{2015MNRAS.454.3816C};
    (33)~\citealp{2017MNRAS.472.3437G};
    (34)~\citealp{2020ApJ...895..118B};
    (35)~\citealp{2016PASA...33...55C};
    (36)~\citealp{2017MNRAS.472.3437G}, note the uncertainty in $\Delta{m}_{15}(B)$ for SN\,2013cs is large, and there is no host galaxy extinction estimate;
    (37)~\citealp{2015MNRAS.452.4307P};
    (38)~\citealp{2016AJ....151..125Z};
    (39)~\citealp{2015MNRAS.452.4307P};
    (40)~\citealp{2019A&A...627A.174H};
    (41)~\citealp{2017MNRAS.464.4476C};
    (42)~This work.}
\end{justify}
    \label{tab:nebref}
\end{table*}

\end{appendix}
\end{document}